\documentclass[numberedappendix,twocolumn]{openjournal}
\usepackage[utf8]{inputenc}
\usepackage{amsmath}
\usepackage[dvipsnames]{xcolor}
\usepackage[colorlinks,allcolors=Blue]{hyperref}

\newcommand{\f}{f_{\mathrm{c}}}
\newcommand{\fzero}{f_{\mathrm{c},0}}
\newcommand{\Rmc}{R_{\mathrm{mc}}}

\DeclareMathOperator\Cov{Cov}

\begin{document}

\title{A new non-parametric method to infer galaxy cluster masses from weak lensing}

\author{Tobias Mistele$^1$\footnote{tobias.mistele@case.edu}, Amel Durakovic$^{2,3}$}
\affiliation{%
$^1$Department of Astronomy, Case Western Reserve University, 10900 Euclid Avenue, Cleveland, Ohio 44106, USA \\
$^2$CEICO — FZU, Institute of Physics of the Czech Academy of Sciences, Na Slovance 1999/2, 182 00 Prague 8, Czechia\\
$^3$Université de Strasbourg, CNRS, Observatoire astronomique de Strasbourg, UMR 7550, 11 Rue de l’Université, 67000 Strasbourg, France
}

\begin{abstract}
We introduce a new, non-parametric method to infer deprojected 3D mass profiles $M(r)$ of galaxy clusters from weak gravitational lensing observations.
The method assumes spherical symmetry and a moderately small convergence, $\kappa \lesssim 1$.
The assumption of spherical symmetry is an important restriction, which is, however, quite common in practice, for example in methods that fit lensing data to an NFW profile.
Moreover, with a mild assumption on the probability distributions of the source redshifts, our method relies on spherical symmetry only at radii larger than the radius $r$ at which the mass $M$ is inferred.
That is, the method may be useful even for clusters with a non-symmetric inner region, since it correctly estimates the enclosed mass beyond the radius where spherical symmetry is restored.
We discuss how to correct, statistically and approximately, for miscentering given that the probability distribution of miscentering offsets is known.
We provide an efficient implementation in Julia code that runs in a few milliseconds per galaxy cluster.
We explicitly demonstrate the method by using data from KiDS DR4 to infer mass profiles for two example clusters, Abell 1835 and Abell 2744, finding results consistent with existing literature.

\end{abstract}

\section{Introduction}

Both strong and weak gravitational lensing are powerful tools to infer the dynamical masses of galaxy clusters \citep[e.g.][]{Bartelmann2001,Umetsu2020,Normann2024}.
A common procedure is to fit a parametric, spherically symmetric mass profile such as an NFW profile \citep{Navarro1996} to lensing observations \citep[e.g.][]{EuclidCollaboration2024,Medezinski2016,Applegate2014}.

Here, we introduce a new method to infer deprojected 3D masses of galaxy clusters from weak gravitational lensing observations.
This method is based on the deprojection formula from \citet{Mistele2023d} which was originally developed for galaxy-galaxy lensing assuming that the convergence $\kappa$ is negligible, $\kappa \ll 1$.
We adapt this method to the case with moderate values of $\kappa$ (which we denote as $\kappa \lesssim 1$) so that it can be applied to galaxy clusters, at least outside the central strong lensing regions. 
This method assumes spherical symmetry but does not assume a specific form of the mass profile, i.e. it is a non-parametric method.
Importantly, it is straightforward to implement and computationally efficient.

The assumption of spherical symmetry is an important restriction.
We note, however, that this assumption is quite common in practice, for example in methods that fit lensing data to a spherically symmetric mass profile \citep[e.g.][]{EuclidCollaboration2024,Medezinski2016,Applegate2014}.
Further, with a mild assumption on the probability distributions of the source redshifts, our method actually requires spherical symmetry only at radii larger than the radius $r$ at which the deprojected mass $M$ is inferred.
Thus, unlike some other methods \citep[e.g.][]{Sommer2022}, our method may be useful even for clusters with a non-symmetric inner region. %
As we will discuss, this property also enables us to correct for miscentering, at least statistically and approximately, with little computational overhead.

We describe our new method and its properties in Sec.~\ref{sec:method}, demonstrate and illustrate it with two explicit example clusters in Sec.~\ref{sec:examples}, and conclude in Sec.~\ref{sec:conclusion}.

\section{The Method}
\label{sec:method}

Our new method is based on a deprojection formula from \citet{Mistele2023d} that allows us to convert excess surface density (ESD) profiles $\Delta \Sigma (R)$ (see below) into deprojected mass profiles $M(r)$ assuming spherical symmetry \citep[see also][]{Mistele2024}.
For galaxy-galaxy lensing, which is what was considered in \citet{Mistele2023d}, $\Delta \Sigma$ is typically straightforward to measure because the convergence $\kappa$ is negligible \citep{Bartelmann2001}.
For galaxy clusters, however, $\kappa$ is not necessarily small.
Thus, an extra step is needed to infer $\Delta \Sigma$ from observations.

\subsection{Main theoretical result}
\label{sec:method:theory}

Weak lensing observations are based on the (complex) ellipticities $\epsilon_s$ of a large number of source galaxies $s$ behind a lens $l$.
These give an estimate of the (complex) reduced shear $g$ around the lens,
\begin{align}
 g = \frac{\gamma}{1 - \kappa} \,,
\end{align}
where $\gamma$ is the (complex) shear and $\kappa$ is the convergence \citep{Bartelmann2001}.
The convergence is given by $\Sigma/\Sigma_{\mathrm{crit},ls}$ where $\Sigma$ is the surface density of the lens and $\Sigma_{\mathrm{crit},ls}$ is the critical surface density,
\begin{align}
\Sigma_{\mathrm{crit},ls}^{-1} = \frac{4 \pi G_{\mathrm{N}}}{c^2} \frac{D(z_l) D(z_l, z_s)}{D(z_s)} \,,
\end{align}
with the Newtonian gravitational constant $G_{\mathrm{N}}$ and the angular diameter distances to the lens, $D(z_l)$, to the source, $D(z_s)$, and between the source and the lens $D(z_l, z_s)$. 

The tangential component $\gamma_+$ of $\gamma$, and therefore also the tangential component $g_+$ of $g$, is closely related to the ESD profile $\Delta \Sigma(R)$ (see below) that we need in order to use the results from \citet{Mistele2023d}.
Indeed, multiplying $\gamma_+$ by the critical surface density and averaging azimuthally gives $\Delta \Sigma$ \citep[e.g.][]{Kaiser1993}.

Thus, we consider the azimuthal average of the tangential reduced shear $g_+$ times the critical surface density,
\begin{align}
\label{eq:G+def}
\begin{split}
G_+ (R)
\equiv \langle \Sigma_{\mathrm{crit},ls} \cdot g_+ \rangle (R)
= \frac{
 \sum_s W_{ls} \cdot \Sigma_{\mathrm{crit},ls} \epsilon_{+,ls}
}{
 \sum_s W_{ls}
}
\,,
\end{split}
\end{align}
where the $W_{ls}$ are weights (see Sec.~\ref{sec:example:data}) and $\epsilon_{+,ls}$ denotes the tangential ellipticity of the source $s$ with respect to the lens $l$.
The sum in Eq.~\eqref{eq:G+def} runs over all sources at a given projected distance $R$ from the lens.

We now assume that the lens is spherically symmetric.
In the following, we will denote its 3D mass density by $\rho(r)$, its surface density (the integral of $\rho$ along the line of sight) by $\Sigma(R)$, and its 3D mass profile by $M(r)$ where $M(r) \equiv 4 \pi \int_0^{r} dr' r'^2 \rho(r')$.
The ESD of the lens is denoted by $\Delta \Sigma$ and is defined in terms of the surface density $\Sigma$,
\begin{align}
 \label{eq:ESD_def}
 \Delta \Sigma(R) \equiv \frac{2}{R^2} \int_0^R dR' R' \Sigma(R') - \Sigma(R) \,.
\end{align}
These definitions of $\Delta \Sigma$ and $M$ apply in spherical symmetry.
In Sec.~\ref{sec:method:nonsymm} we relax the assumption of spherical symmetry and give definitions that apply more generally.

Assuming spherical symmetry, $G_+ (R)$ is related to the surface density $\Sigma(R)$ and the ESD profile $\Delta \Sigma(R)$ of the lens by \citep{Seitz1997,Umetsu2020},
\begin{align}
 \label{eq:GSigmarelation}
 G_+ (R) \simeq \frac{\Delta \Sigma(R)}{1 - \f(R) \Sigma(R)} \,,
\end{align}
where the "$\simeq$" indicates that this relation is valid only as long as $\f \cdot \Sigma$ is not too large \citep{Seitz1997} and $\f$ is the average inverse critical surface density,
\begin{align}
 \f(R) \equiv \langle \Sigma_{\mathrm{crit},ls}^{-1} \rangle (R) \,.
\end{align}
As in Eq.~\eqref{eq:G+def}, the average runs over the sources at a projected distance $R$ and uses weights $W_{ls}$.
In the following, we assume that $\f \cdot \Sigma$ is indeed not too large so that Eq.~\eqref{eq:GSigmarelation} is valid.
We denote this condition by $\f(R) \Sigma(R) \lesssim 1$ which is roughly equivalent to
\begin{align}
 \label{eq:fSigma_condition}
 \kappa = \Sigma/\Sigma_{\mathrm{crit}} \lesssim 1 \,.
\end{align}

Both $G_+$ and $\f$ are observational quantities that can be measured directly.
Our goal is to infer the deprojected mass of the lens, given these two quantities.
To achieve this, we will use the relation Eq.~\eqref{eq:GSigmarelation} as well as previous results from \citet{Mistele2023d}.
We also make the mild additional assumption that the mass density $\rho$ of the lens asymptotically falls off faster than $1/r$.

Our strategy is as follows.
First, we eliminate $\Sigma$ from Eq.~\eqref{eq:GSigmarelation} by using the following relation between $\Sigma$ and $\Delta \Sigma$ which follows from Eq.~\eqref{eq:ESD_def} \citep{Mistele2023d},
\begin{align}
 \label{eq:ESD_SD_relation}
 \Sigma(R) = - \Delta \Sigma(R) + \int_R^\infty d R' \frac{2 \Delta \Sigma(R')}{R'} \,.
\end{align}
This gives a relation between the observational quantities $G_+$ and $\f$ on the one hand and $\Delta \Sigma$ on the other hand, eliminating the dependence on $\Sigma$.
The next step is to solve this relation for $\Delta \Sigma$ which gives $\Delta \Sigma$ as a function of $G_+$ and $\f$.
Finally, we use the deprojection formula from \citet{Mistele2023d} which, given $\Delta \Sigma$, allows us to calculate the dynamical deprojected mass $M(r)$ within a spherical radius $r$.

Details of this derivation are given in Appendix~\ref{sec:appendix:derivation}.
The final result is
\begin{align}
 \label{EQ:M_FROM_ESD}
 \frac{M(r)}{r^2} = 4 \int_0^{\pi/2} d \theta \, \Delta \Sigma \left( \frac{r}{\sin \theta}\right) \,,
\end{align}
which is the deprojection formula from \citet{Mistele2023d}, with the following specific $\Delta \Sigma$,
\begin{widetext}
\begin{align}
 \label{EQ:ESD_FROM_GF} %
 \Delta \Sigma (R)
 =
 \frac{1}{\f(R)} 
 \frac{G_+ \f(R)}{1 - G_+ \f(R)}
 \left[
 1
 - \int_R^\infty dR'' \frac{\f(R)}{\f(R'')} \frac{2}{R''} \frac{G_+\f(R'')}{1 - G_+\f(R'')}
                      \exp\left(
                        -\int_R^{R''} d R' \frac{2}{R'} \frac{G_+\f(R')}{1 - G_+\f(R')}
                      \right)
  \right]
  \,,
\end{align}
\end{widetext}
where we introduced the shorthand notation $G_+ \f(R) \equiv G_+(R) \cdot \f(R)$.
This allows us to calculate the dynamical mass $M(r)$ from the observational quantities $G_+$ and $\f$.
Thus, Eq.~\eqref{EQ:ESD_FROM_GF} together with Eq.~\eqref{EQ:M_FROM_ESD} is the core theoretical result of this work.

In practice, the average inverse critical surface density $\f(R)$ is often approximately constant as a function of $R$, i.e. $\f(R) \approx \mathrm{const}$.
In this case, Eq.~\eqref{EQ:ESD_FROM_GF} simplifies,
\begin{widetext}
\begin{align}
 \label{eq:ESD_from_Gf_fconst}
 \left. \Delta \Sigma (R) \right|_{\f = \mathrm{const}} = 
 \frac{1}{\f}
 \frac{G_+ \f(R)}{1 - G_+\f(R)} \exp\left(
  - \int_R^\infty dR' \frac{2}{R'} \frac{G_+ \f(R')}{1 - G_+ \f(R')}
 \right) \,.
\end{align}
\end{widetext}
In the limit $\kappa \ll 1$, the relation between $\Delta \Sigma$ and the observational quantities $G_+$ and $\f$ becomes even simpler, namely $\Delta \Sigma (R) \approx G_+ (R)$.
This last relation is what the galaxy-galaxy lensing analysis in \citet{Mistele2023d} is based on.

\begin{figure*}
\begin{center}
\includegraphics[width=2.1\columnwidth]{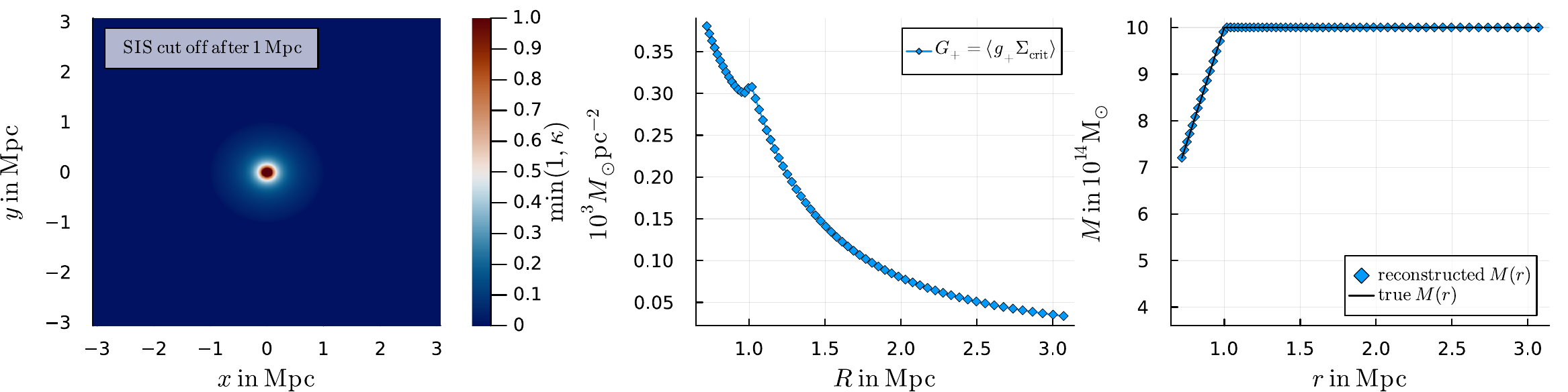}
\includegraphics[width=2.1\columnwidth]{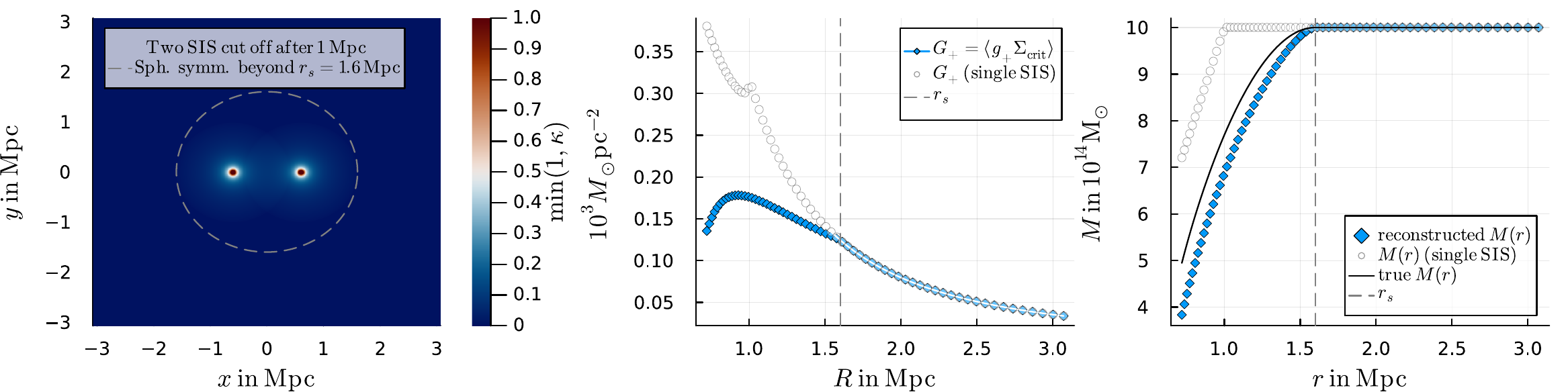}
\end{center}
\caption{
 Illustration of our method for lenses with a non-symmetric inner region, demonstrating that it infers the correct mass $M(r)$ at radii $r$ larger than the radius $r_s$ beyond which the lens's 3D density is spherically symmetric.
 Lenses are assumed to be at $z_l = 0.3$.
 Sources are assumed to be at $z_s = 0.9$ and $z_s=1.2$, following the same spatial distribution at both redshifts so that $f_c = \langle \Sigma_{\mathrm{crit}}^{-1} \rangle$ is constant.
 We assume the same cosmology as in Sec.~\ref{sec:example:data}.
 \emph{Top row}: For reference, an SIS whose 3D density profile is sharply cut off at a spherical distance $r=1\,\mathrm{Mpc}$.
 Its total mass is $10^{15}\,M_\odot$.
 \emph{Left}: The convergence $\kappa$, in this plot taken to be $\Sigma \cdot \langle \Sigma_{\mathrm{crit}} ^{-1} \rangle$.
 \emph{Middle}: The azimuthally averaged tangential reduced shear $G_+ = \langle g_+ \Sigma_{\mathrm{crit}} \rangle$.
 \emph{Right}: The mass $M(r)$ reconstructed from $G_+$ and $f_c$ using our deprojection formulas Eq.~\eqref{eq:ESD_from_Gf_fconst} and Eq.~\eqref{EQ:M_FROM_ESD} (blue symbols) and the true mass profile $M(r)$ (solid black line).
 We extrapolate $G_+$ assuming it asymptotically falls off as $1/R^2$, corresponding to a point mass. 
 Since this SIS is spherically symmetric at all radii, our deprojection method recovers the correct mass $M(r)$ at all radii.
 \emph{Bottom row}: Two SIS profiles that are cut off as in the top row, but offset by $\pm 0.6\,\mathrm{Mpc}$ along the $x$-axis and each with a total mass of $5 \cdot 10^{14}\,M_\odot$ (half that of the SIS from the top row).
 The 3D density profile of the combined two-SIS system is spherically symmetric beyond $r_s = 1.6\,\mathrm{Mpc}$ but not at smaller radii.
 The three panels are as in the top row.
 The bottom middle panel shows that $G_+$ at $R > r_s$ is the same as for the single SIS from the top row.
 The bottom right panel shows that our deprojection method recovers the correct mass at $r > r_s$, where the 3D density is spherically symmetric, despite the existence of a non-symmetric inner region.
}
\label{fig:nonsymmetricdemo}
\end{figure*}

The above assumes that an estimate of $\Sigma_{\mathrm{crit},ls}$ is available for each individual source $s$.
If this is not the case and only ensemble information is available, we can still use Eq.~\eqref{EQ:M_FROM_ESD} and Eq.~\eqref{EQ:ESD_FROM_GF} to estimate the deprojected mass profile $M(r)$, if we adapt the definitions of $G_+$ and $f_c$,
\begin{subequations}
 \label{eq:G_and_fc_def_ensemble}
\begin{alignat}{2}
 &G_+ = \langle \Sigma_{\mathrm{crit},ls} \, g_+ \rangle &&\quad \to \quad G_+ = \frac{\langle g_+ \rangle}{\langle \Sigma_{\mathrm{crit},ls}^{-1} \rangle}\,, \\
 &f_c = \langle \Sigma_{\mathrm{crit},ls}^{-1} \rangle &&\quad \to \quad f_c = \frac{\langle \Sigma_{\mathrm{crit},ls}^{-2} \rangle}{\langle \Sigma_{\mathrm{crit},ls}^{-1} \rangle} \,.
\end{alignat}
\end{subequations}
This can be seen by retracing the above derivation with Eq.~\eqref{eq:GSigmarelation} replaced by the analogous relation \citep{Umetsu2020,Applegate2014,Seitz1997}
\begin{align}
 \frac{\langle g_+ \rangle}{\langle \Sigma_{\mathrm{crit},ls}^{-1} \rangle}
 \simeq
 \frac{
  \Delta \Sigma
 }{
  1 - \frac{\langle \Sigma_{\mathrm{crit},ls}^{-2} \rangle}{\langle \Sigma_{\mathrm{crit},ls}^{-1} \rangle} \, \Sigma
 } \,,
\end{align}
which also holds as long as $\kappa$ is not too large.
In practice, if individual source redshifts are not available, a separate estimate of $f_c$ for each radius $R$ is likely also not available.
In this case, one may as well use Eq.~\eqref{eq:ESD_from_Gf_fconst} instead of Eq.~\eqref{EQ:ESD_FROM_GF} to simplify the calculation.

Mathematically, both Eq.~\eqref{EQ:ESD_FROM_GF} and Eq.~\eqref{eq:ESD_from_Gf_fconst} are only valid as long as $G_+\f(R)$ remains smaller than $1$.
This roughly corresponds to the condition that the reduced tangential shear $g_+$ is smaller than $1$, $g_+ < 1$.
In fact, this condition is implicit in weak-lensing results we build on such as Eq.~\eqref{eq:G+def}.
Thus, it is not an important additional restriction; it is mostly just a reminder that we are working within the weak-lensing regime.

From Eq.~\eqref{EQ:ESD_FROM_GF} and Eq.~\eqref{EQ:M_FROM_ESD} we see that the inferred mass $M(r)$ at a radius $r$ depends on observations only from radii $R$ larger than $r$.
This is useful in case no data from small radii is available.
But perhaps more importantly, it also is the basis for another, potentially more generally useful property of our method.

\subsection{Spherical symmetry not required at small radii}
\label{sec:method:nonsymm}

This property is that our method works even for clusters with a non-symmetric inner region.
That is, the results from the previous subsection hold even when the mass density $\rho$ of the lens is spherically symmetric only beyond some spherical radius $r_s$.
More precisely, Eqs.~\eqref{EQ:ESD_FROM_GF} and \eqref{EQ:M_FROM_ESD} still give the correct mass $M(r)$ at radii $r$ larger than $r_s$ even when the lens is not spherically symmetric at radii smaller than $r_s$.
In practice, this may be relevant, for example, for late-stage mergers or clusters with a complex baryonic mass distribution.

Here, by the mass $M(r)$ we mean the mass within a sphere centered on the origin of the coordinate system,
\begin{align}
 M(r) \equiv \int_{|\vec{x}| < r} d^3 \vec{x} \, \rho(\vec{x}) \,,
\end{align}
where $\rho(\vec{x}$) is the 3D mass density.
By the assumption that $\rho$ is spherically symmetric beyond some radius $r_s$ we mean that, in this coordinate system, $\rho(\vec{x}) = \rho(|\vec{x}|)$ for $|\vec{x}| \geq r_s$.

As we will see below, to rigorously derive this result, we make use of one additional assumption which is that the redshifts of the source galaxies follow probability distributions that do not depend on the azimuth (but may depend on projected radius $R$).
That is, we assume that, at each position $(R \cos \varphi, R \sin \varphi)$, the source redshifts are drawn from a probability distribution $p_{z_s}(z_s|R)$ that does not depend on $\varphi$.
This disallows a lopsidedness in the source redshift population but does allow a radial variation, for example due to obscuration towards the cluster center.
We expect that this is often a reasonable assumption in practice.

\begin{figure*}
\begin{center}
\includegraphics[width=2.1\columnwidth]{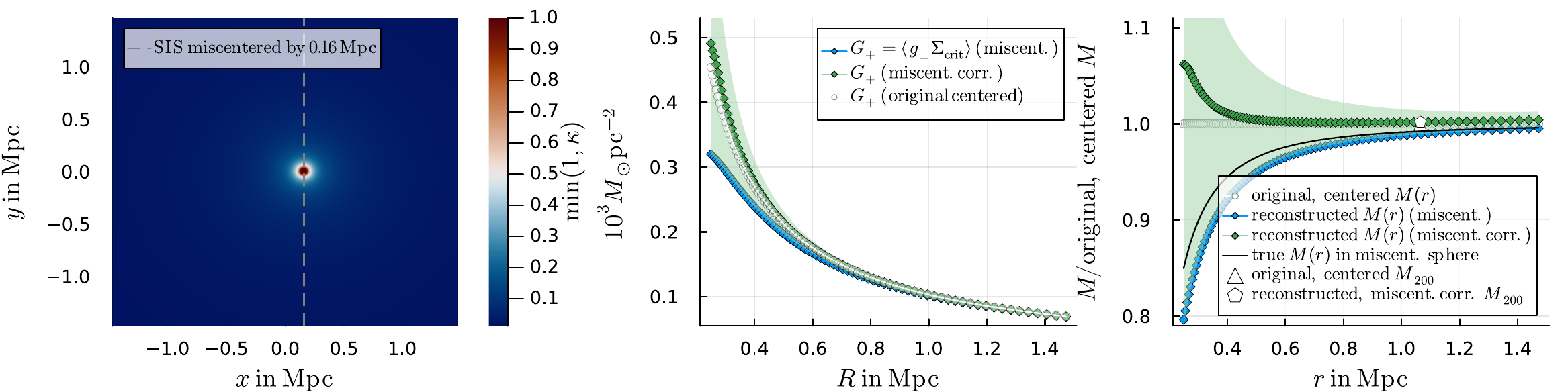}
\end{center}
\caption{
 Illustration of how miscentering may, approximately, be taken into account.
 We consider an SIS at $z_l=1.0$ miscentered by $160\,\mathrm{kpc}$.
 The original, centered SIS has a mass of $4 \cdot 10^{14}\,M_\odot$ within a spherical radius of $1\,\mathrm{Mpc}$.
 All sources are assumed to be at $z_s=1.7$.
 We assume the same cosmology as in Sec.~\ref{sec:example:data}.
 \emph{Left}: The convergence $\kappa$.
 \emph{Middle}: The azimuthally averaged tangential reduced shear $G_+ = \langle g_+ \Sigma_{\mathrm{crit}} \rangle$ for the miscentered SIS (blue symbols), for the original, centered SIS (white symbols), and for the miscentered SIS corrected for miscentering using Eq.~\eqref{EQ:MISCENTERCORRECT} (green symbols).
 The miscentering correction brings the $G_+$ of the miscentered SIS close to that of the original, centered SIS.
 The light green band illustrates how an uncertainty in $\Rmc^2$, here taken to be $(160\,\mathrm{kpc})^2$ \citep[corresponding to a Rayleigh distribution,][]{Johnston2007}, would induce uncertainties in the miscentering-corrected $G_+$ even if observations are otherwise noiseless.
 \emph{Right}: The mass $M(r)$ reconstructed from $G_+$ and $f_c = \langle \Sigma_{\mathrm{crit}}^{-1} \rangle$ using our deprojection formulas Eq.~\eqref{eq:ESD_from_Gf_fconst} and Eq.~\eqref{EQ:M_FROM_ESD}, relative to the mass of the original, centered SIS.
 The mass reconstructed from the miscentered SIS (blue symbols) is quite close to the true mass within a sphere with radius $r$ centered on the \emph{origin} (solid black line).
 It is, however, not as close to the mass within a sphere with radius $r$ centered on the \emph{actual center} of the SIS (white symbols), which is what one likely cares about in practice.
 Applying the miscentering correction from Eq.~\eqref{EQ:MISCENTERCORRECT} helps (green symbols).
 The light green band illustrates how an uncertainty in $\Rmc^2$ of $(160\,\mathrm{kpc})^2$ would induce uncertainties in the reconstructed, miscentering-corrected mass even for otherwise noiseless observations.
 This green band reaches precisely to the un-corrected reconstructed mass because, within our approximation, the miscentering-corrected mass is linear in $\Rmc^2$.
}
\label{fig:miscenteringdemo}
\end{figure*}

Fig.~\ref{fig:nonsymmetricdemo} illustrates this property, that our method allows for non-symmetry at small radii, for a simple toy example, consisting of two SIS (singular isothermal sphere) profiles at $z_l = 0.3$ whose 3D densities are sharply cut off after $1\,\mathrm{Mpc}$ and which are off-centered along the $x$-axis by $\pm 0.6\,\mathrm{Mpc}$.
The left panel of Fig.~\ref{fig:nonsymmetricdemo} shows $\Sigma \cdot \langle \Sigma_{\mathrm{crit}}^{-1} \rangle$ to illustrate the setup.
The middle panel illustrates the main input to our deprojection method $G_+ = \langle g_+ \Sigma_{\mathrm{crit}} \rangle$.
We consider sources to be at $z_s = 0.9$ and $z_s = 1.2$, with the same spatial distribution at both values of $z_s$ so that $\f = \langle \Sigma_{\mathrm{crit}}^{-1} \rangle$ does not depend on $R$ for simplicity.
To obtain $G_+$, we first calculate the projected surface density $\Sigma$, then we calculate the lensing potential, then, for each value of $z_s$, we calculate the tangential shear $\gamma_+$, the convergence $\kappa$, and the reduced tangential shear $\gamma_+/(1 - \kappa)$, and finally we average azimuthally and over the different $z_s$.
The right panel of Fig.~\ref{fig:miscenteringdemo} illustrates the result of applying our deprojection formulas Eq.~\eqref{EQ:ESD_FROM_GF} and Eq.~\eqref{EQ:M_FROM_ESD} to these simple mock values of $G_+$ and $\f$.

The 3D density of the combined two-cut-off-SIS system from Fig.~\ref{fig:nonsymmetricdemo} is exactly spherically symmetric beyond $r_s = 1.6\,\mathrm{Mpc}$, but is not spherically symmetric at smaller radii.
Thus, as shown in Fig.~\ref{fig:nonsymmetricdemo}, right, our method does not infer the correct mass at small radii.
It does, however, infer the correct mass beyond $r_s$, despite the existence of a non-symmetric inner region.

This may be surprising given that the derivation of our deprojection formulas assumes spherical symmetry \emph{at all radii} (this is the case for both the first step of obtaining $\Delta \Sigma$ from $G_+$ and $\f$ using Eq.~\eqref{EQ:ESD_FROM_GF} and the second step of obtaining $M$ from $\Delta \Sigma$ using Eq.~\eqref{EQ:M_FROM_ESD}).
In the rest of this subsection we show why, despite this assumption of spherical symmetry at all radii, the resulting formulas work, at $r > r_s$, even in the case where the lens is spherically symmetric only at $r > r_s$.

To this end, consider a lens that is spherically symmetric only at $r > r_s$.
In this case, the relation Eq.~\eqref{eq:GSigmarelation} between $G_+$,  $\f$, $\Delta \Sigma$, and $\Sigma$ still holds at $R > r_s$.
We just need to replace $\Sigma$ by its azimuthal average $\langle \Sigma \rangle$ in the definition Eq.~\eqref{eq:ESD_def} of $\Delta \Sigma$.
That is,
\begin{align}
\label{EQ:GSIGMARELATION_NONSYMM}
G_+ (R) \approx \frac{\Delta \Sigma (R)}{1 - \f(R) \, \Sigma (R) } \quad (\mathrm{for}\; R > r_s) \,,
\end{align}
with $\Delta \Sigma$ defined as
\begin{align}
 \label{eq:ESD_def_nonsymm_maintext}
 \Delta \Sigma(R) \equiv \frac{2}{R^2} \int_0^R dR' R' \langle \Sigma \rangle(R') - \langle \Sigma \rangle(R) \,.
\end{align}
We derive this result in Appendix~\ref{sec:appendix:nonsymm}.
This derivation makes use of the assumption that the source redshifts follow probability distributions $p_{z_s}(z_s|R)$ that are independent of the azimuth $\varphi$.
It also makes use of the fact that, since the density $\rho$ is spherically symmetric for $r > r_s$, $\Sigma$ is cylindrically symmetric at $R > r_s$.
This is also why the $\Sigma(R)$ in the denominator of Eq.~\eqref{EQ:GSIGMARELATION_NONSYMM} is well-defined.

The definition Eq.~\eqref{eq:ESD_def_nonsymm_maintext} of $\Delta \Sigma$ can alternatively be written as
\begin{align}
 \label{eq:ESD_def_M2d}
 \Delta \Sigma (R) = \frac{M_{2D}(R)}{\pi R^2} - \langle \Sigma \rangle (R) \,,
\end{align}
where $M_{2D} (R)$ is the mass within a cylinder with radius $R$ that is oriented along the line of sight and centered on the origin.
Thus, the value of $\Delta \Sigma$ at a projected radius $R$ is sensitive only to the total mass within that cylinder and to the surface density $\langle \Sigma \rangle$ at $R$.
How the mass inside that cylinder is distributed does not affect its value.

Equation~\eqref{EQ:GSIGMARELATION_NONSYMM} then implies that the same property also holds for $G_+(R)$ at $R > r_s$.
In particular, the value of $G_+$ at $R > r_s$ is independent of how the mass within the projected distance $R$ is distributed.
This property is due to the azimuthal average in $G_+ = \langle g_+ \Sigma_{\mathrm{crit}} \rangle$.
Had we considered $g_+ \Sigma_{\mathrm{crit}}$ without the azimuthal average, we would not have this property.

Consider then a second lens with the same mass distribution, lens redshift, and background source redshifts as the original lens, except with one change:
All the mass within a spherical radius $r_s$ around this second lens is redistributed to be spherically symmetric.
For example, the mass within $r_s$ may be collapsed into a single point at the origin.
As an illustration, consider going from the bottom row of Fig.~\ref{fig:nonsymmetricdemo} to the top row.
By construction, these two lenses produce the same values for $\f = \langle \Sigma_{\mathrm{crit}}^{-1} \rangle$.
Further, these two lenses differ only in their mass distribution at projected radii $R$ smaller than $r_s$.
Thus, as we just argued based on Eq.~\eqref{EQ:GSIGMARELATION_NONSYMM} and Eq.~\eqref{eq:ESD_def_M2d}, these two lenses also produce the same $G_+ (R)$ at $R > r_s$ (for example, compare the blue symbols in the two middle panels of Fig.~\ref{fig:nonsymmetricdemo} at $R > 1.6\,\mathrm{Mpc}$).

Finally, consider \emph{spherical} radii $r > r_s$.
When applied to such radii $r$, the deprojection formulas Eq.~\eqref{EQ:ESD_FROM_GF} and Eq.~\eqref{EQ:M_FROM_ESD} make use of the observables $G_+$ and $\f$ only at projected radii \emph{larger} than $r_s$, i.e. $R > r_s$.
Thus, if we apply these deprojection formulas to spherical radii $r > r_s$, we will get the same result for both the original lens and the second lens that has all the mass within $r_s$ redistributed to be spherically symmetric (for example, compare the blue symbols in the two right panels of Fig.~\ref{fig:nonsymmetricdemo} at $r > 1.6\,\mathrm{Mpc}$).

For the second lens, with all the mass within $r_s$ redistributed to be spherically symmetric, we know that our deprojection formulas infer the correct mass profile $M(r)$ because that lens is spherically symmetric everywhere (for example, compare the solid black line and the blue symbols in the top right panel of Fig.~\ref{fig:nonsymmetricdemo}).
Further, by construction the two lenses have the same mass profile $M(r)$ beyond $r_s$.
Thus, it follows that, for $r > r_s$, our deprojection formulas infer the correct mass profile $M(r)$ also for the original lens with the non-symmetric inner region (for example, compare the solid black line and the blue symbols in the bottom right panel of Fig.~\ref{fig:nonsymmetricdemo}).
This completes the argument.

It remains to be seen how well this theoretical result works in practice where the assumptions we make, although reasonable, are never exactly satisfied.
For example, unlike in Fig.~\ref{fig:nonsymmetricdemo}, in realistic scenarios there may not be a radius beyond which the 3D density $\rho$ becomes exactly spherically symmetric.
It is plausible that, at large radii, such systems are sufficiently spherically symmetric for our deprojection method to produce a good approximation of the true mass profile.
But ultimately the practical usefulness of the above result must be further validated in future work, for example using $N$-body simulations.

\subsection{Practical considerations}
\label{sec:method:practice}

To use Eq.~\eqref{EQ:M_FROM_ESD} and Eq.~\eqref{EQ:ESD_FROM_GF} in practice requires a few additional considerations.
For example, the integrals in Eq.~\eqref{EQ:ESD_FROM_GF} extend to $\infty$ and therefore require $G_+ (R)$ and $\f (R)$ at arbitrarily large radii $R$.
In practice, however, these are measured only up to some finite maximum radius $R_{\mathrm{max}}$.
To deal with this, we extrapolate $G_+(R)$ and $\f(R)$ beyond $R_{\mathrm{max}}$ using a simple power law for $G_+$ and a constant for $\f$,
\begin{subequations}
 \label{eq:extrapolate}
 \begin{align}
 G_+(R > R_{\mathrm{max}}) &\equiv G_+(R_{\mathrm{max}}) \cdot \left(\frac{R_{\mathrm{max}}}{R}\right)^n \,, \\
 \f(R > R_{\mathrm{max}}) &\equiv \f(R_{\mathrm{max}}) \,.
 \end{align}
\end{subequations}
This extrapolation is an important systematic uncertainty.
However, in practice, it becomes important only close to the last measured data point at $R = R_{\mathrm{max}}$.
Over most of the radial range, the extrapolation has little impact on the inferred mass $M(r)$ (see Sec.~\ref{sec:examples} below).
This is because most of the signal in the inferred $M(r)$ comes from $G_+(R)$ and $\f(R)$ at $R \sim r$.

For our explicit examples below, we extrapolate assuming $n=1$ which corresponds to an SIS, unless otherwise stated.
We take the uncertainty in this choice into account as a systematic error.
In particular, we adopt the difference between extrapolation with $n=1/2$ and $n=2$ as a systematic error.
Schematically,
\begin{align}
\left. \sigma_{M (r)}\right|_{\mathrm{syst}}^{\mathrm{extrapolate}} \equiv \left| \left.M(r)\right|^{n=2} -  \left.M(r)\right|^{n=1/2} \right| \,.
\end{align} 
For comparison, $n=2$ corresponds to $G_+$ asymptotically dropping off as for a point mass.
For $n=1/2$, $G_+$ drops off slower than for a singular isothermal sphere.
These two extreme cases likely bracket the true behavior of $G_+$.

Similarly, $G_+$ and $\f$ are measured in discrete radial bins but the continuous integrals in Eq.~\eqref{EQ:ESD_FROM_GF} require them at all radii.
Thus, we interpolate between the discrete radial bins.
For our explicit examples below we use linear interpolation and we adopt the difference between linear and quadratic interpolation as an estimate of the systematic error due to interpolation.
Schematically,
\begin{align}
\left. \sigma_{M (r)}\right|_{\mathrm{syst}}^{\mathrm{interpolate}} \equiv \left| \left.M(r)\right|^{\mathrm{quadratic}} -  \left.M(r)\right|^{\mathrm{linear}} \right| \,.
\end{align}
We interpolate linearly (or quadratically) in $R$.
Another reasonable option is to interpolate in $\ln R$.
For our explicit examples below, this choice does not make a significant difference.

For the statistical uncertainties and covariances, we use linear error propagation to convert uncertainties on $G_+$ into uncertainties and covariances on the inferred mass $M$.
Linear error propagation requires calculating the Jacobian of the inferred mass $M(r)$ understood as a function of the measured values $G_+(R_i)$ where $R_i$ denote the discrete radii where $G_+$ is measured.
This calculation is cumbersome to do by hand.
To avoid this, we wrote differentiable numerical Julia code to evaluate Eq.~\eqref{EQ:ESD_FROM_GF} and Eq.~\eqref{EQ:M_FROM_ESD}.
Differentiable here means that the Julia package `ForwardDiff.jl` \citep{Revels2016} can automatically calculate the needed Jacobians for us.
Linear error propagation then reduces to a simple matrix multiplication.
Our code is publicly available\footnote{\url{https://github.com/tmistele/SphericalClusterMass.jl}} and takes only a few milliseconds per galaxy cluster to run on a standard personal computer.
In Appendix~\ref{sec:appendix:numerical}, we discuss how we efficiently evaluate the integrals in Eq.~\eqref{EQ:M_FROM_ESD} and Eq.~\eqref{EQ:ESD_FROM_GF}.

\subsection{Miscentering}
\label{sec:method:miscentering}

Another practical concern is miscentering which may be particularly relevant for lower-mass clusters \citep[e.g.][]{Rozo2011}.
How to take miscentering into account will depend on what precisely one is interested in.

If one requires precise masses $M(r)$ at small radii $r$ (comparable to or smaller than the miscentering offset $R_{\mathrm{mc}}$) then a brute force calculation of some kind may be necessary.
For example, one may redo the deprojection a large number of times, each time re-centering according to an empirical probability distribution $p_{\mathrm{mc}}$ of miscentering offsets.
This would allow to statistically infer the expected mass $M(r)$ with respect to a given miscentering probability distribution $p_{\mathrm{mc}}$.
Such procedures have a significant computational cost, but might not be prohibitive given that the deprojection runs fast.

In any case, there is often a better way with very little computational overhead.
Indeed, in practice one is often mainly interested in $M(r)$ at radii $r$ that are much larger than typical miscentering offsets $\Rmc$, for example when one wants to measure the virial mass $M_{200}$.

In these cases, it is beneficial that our method works even for clusters with a non-symmetric inner region (Sec.~\ref{sec:method:nonsymm}).
This is because miscentering essentially induces a non-symmetry at small radii but leaves the profile approximately symmetric at large radii.
This is illustrated in Fig.~\ref{fig:miscenteringdemo} for a miscentered SIS profile:
At radii $r$ larger than the offset $\Rmc$, our deprojection formulas work quite well for the purpose of recovering the true mass within a sphere with radius $r$ centered at the origin (compare the solid black line and the blue symbols in the right panel of Fig.~\ref{fig:miscenteringdemo}).

Of course, for a miscentered halo one is usually \emph{not} actually interested in how well one can recover the mass within a sphere centered on the \emph{origin} but how well one can recover the mass within a sphere centered on the \emph{actual, original center} of the halo.
Fig.~\ref{fig:miscenteringdemo} shows that our deprojection method also recovers this original, centered mass profile reasonably well at large radii (compare the blue and white symbols in the right panel), but not as well as the mass within the mis-centered sphere.

We will now discuss how to improve on this by systematically expanding in the parameter
\begin{align}
 \epsilon \equiv \frac{\Rmc}{R} \,,
\end{align}
which is typically small at the large radii relevant to measure, for example, $M_{200}$.
As in Sec.~\ref{sec:method:nonsymm}, we will assume that the source redshifts follow probability distributions $p_{z_s}(z_s|R)$ that do not depend on the azimuth.\footnote{
 To our order of approximation, it does not matter whether this assumption is made in the miscentered or the original, centered frame.
 As explained in Appendix~\ref{sec:appendix:miscentering}, the difference is of order $\kappa \epsilon^2$, which we neglect in the following.
}

As we show in Appendix~\ref{sec:appendix:miscentering}, one can take miscentering into account by a simple pre-processing step of $G_+(R)$ before feeding it into our deprojection formulas Eq.~\eqref{EQ:ESD_FROM_GF} and Eq.~\eqref{EQ:M_FROM_ESD}.
In particular, we need to replace
\begin{align}
 \label{EQ:MISCENTERCORRECT}
 G_+(R) \to G_+(R) + \frac14 \left(\frac{\Rmc}{R}\right)^2 \Delta_{\mathrm{mc}} (R)\,,
\end{align}
where
\begin{align}
  \Delta_{\mathrm{mc}}(R) \equiv 4 G_+ (R) - R \partial_R G_+(R) - R^2 \partial_R^2 G_+(R).
\end{align}
This formula includes all corrections due to miscentering up to and including order $\epsilon^2$ except those of order $\kappa \epsilon^2$.
That is, we treat $\kappa$ as small so that corrections of order $\kappa \epsilon^2$ can be neglected compared to the leading-order correction of order $\kappa^0 \epsilon^2$.

This correction works well at radii that are large compared to $\Rmc$, as illustrated in the middle and right panels of Fig.~\ref{fig:miscenteringdemo} (compare the green and white symbols):
Using Eq.~\eqref{EQ:MISCENTERCORRECT} brings the reconstructed mass profile very close to the original, centered mass profile.
In this toy example, the original, centered $M_{200}$ and the $M_{200}$ inferred from the reconstructed, miscentering-corrected mass profile differ by less than $3$ permill.

Of course, in practice one usually does not know the true miscentering offset $\Rmc$.
So, unlike in our toy example from Fig.~\ref{fig:miscenteringdemo}, one cannot precisely correct for the miscentering.
One may, however, know the probability distribution $p_{\mathrm{mc}}(\Rmc)$ of miscentering offsets.
In this case, one can still correct for miscentering statistically.

For example, one can still calculate the expected mass $\langle M(r) \rangle_{\mathrm{mc}}$, where $\langle \dots \rangle_{\mathrm{mc}}$ denotes averaging with respect to $p_{\mathrm{mc}}$,
\begin{align}
\label{eq:pmcintegral}
\langle M(r) \rangle_{\mathrm{mc}} = \int_0^\infty d\Rmc \, p_{\mathrm{mc}}(\Rmc) M(r|\Rmc^2) \,,
\end{align}
and where $M(r|\Rmc^2)$ denotes the mass obtained by first correcting $G_+$ for miscentering using Eq.~\eqref{EQ:MISCENTERCORRECT} and then applying the deprojection procedure from Sec.~\ref{sec:method:theory}.

In fact, within our order of approximation, the integral Eq.~\eqref{eq:pmcintegral} is straightforward to evaluate.
This is because, to order $\epsilon^2 = (\Rmc/R)^2$, the deprojected mass $M(r|\Rmc^2)$ is linear in $\Rmc^2$.
Schematically,
\begin{multline}
 \label{eq:MRmc_linearity}
 M(r|\Rmc^2) = M(r|\Rmc^2 = 0) \\
  +  \Rmc^2 \cdot (\mathrm{corrections}) + (\mathrm{higher\;orders}) \,.
\end{multline}
This follows from the fact that the leading order correction to the inputs of our deprojection formulas, specifically $G_+$, is of order $\epsilon^2$ (see Eq.~\eqref{EQ:MISCENTERCORRECT}).

Thus, the effect of the $\Rmc$ integral in Eq.~\eqref{eq:pmcintegral} is simply to replace $\Rmc^2$ by its expectation value $\langle \Rmc^2 \rangle_{\mathrm{mc}}$.
In other words, the expected deprojected mass $\langle M(r) \rangle_{\mathrm{mc}}$ can be obtained by first correcting $G_+$ for miscentering using Eq.~\eqref{EQ:MISCENTERCORRECT}, using the expected $\langle \Rmc^2 \rangle_{\mathrm{mc}}$ instead of the (unknown) true $\Rmc^2$, and then following the deprojection procedure as described in Sec.~\ref{sec:method:theory}.
Symbolically,
\begin{align}
 \langle M(r) \rangle_{\mathrm{mc}} = M(r|\Rmc^2 = \langle \Rmc^2 \rangle_{\mathrm{mc}}) \,.
\end{align}

This procedure is implemented in the provided Julia code.
This code also automatically corrects the covariance matrix of $G_+$ according to Eq.~\eqref{EQ:MISCENTERCORRECT} (also propagating the uncertainty of $\langle \Rmc^2 \rangle_{\mathrm{mc}}$ which enters this equation).
Covariances and uncertainties are again propagated using automatic differentiation.
To illustrate, the green band in the right panel of Fig.~\ref{fig:miscenteringdemo} shows how an uncertainty on $\Rmc^2$ would induce uncertainties in the deprojected mass even for otherwise noiseless observations.

Similar techniques and results may also be useful when one is interested in quantities other than the expected mass $\langle M(r) \rangle_{\mathrm{mc}}$ (for example when deriving statistical constraints on cosmological parameters).
This will be investigated in future work.
Moreover, we note that the prescription Eq.~\eqref{EQ:MISCENTERCORRECT} of correcting the observed reduced tangential shear is not tied to our deprojection method.
So it might be a useful tool in other approaches as well.

In any case, the explicit example from Fig.~\ref{fig:miscenteringdemo} is of course only a simple, noiseless toy model.
In future work, we will investigate how well this procedure works in a more realistic setting, for example using $N$-body simulations.
It also remains to work out the formalism to implement this procedure as part of a stacked analysis.

\section{Explicit Examples}
\label{sec:examples}

We now demonstrate our method explicitly using two galaxy clusters listed in the LC2 catalog \citep{Sereno2015}, Abell 1835 and Abell 2744.
Abell 2744 is a merging system for which our assumption of spherical symmetry is likely not a good approximation.
It is, however, still useful to compare the results of our method to the results of other methods that also assume spherical symmetry such as the single-halo model from \citet{Medezinski2016}.

We use weak-lensing data from KiDS DR4 \citep{Kuijken2019,Giblin2021}.
We select Abell 1835 and Abell 2744 as the two clusters from LC2 with the highest source number density in KiDS (with the source number density being measured in $\mathrm{arcmin}^{-2}$ and taking into account sources within a projected distance of $3\,\mathrm{Mpc}$).
These clusters are among the 5\% most massive clusters in LC2 (according to the $M(<1\,\mathrm{Mpc})$ listed there) which, together with their relatively high source number densities, ensures a reasonable signal-to-noise ratio.
As we will see, our new method gives results compatible with previous mass measurements from \citet{Applegate2014} and \citet{Medezinski2016}.

\subsection{Data}
\label{sec:example:data}

We mostly follow the procedure of \citet{Mistele2023d} and \citet{Brouwer2021}.
In particular, we use ellipticities from the KiDS-1000 SOM-gold source galaxy catalog \citep{Kuijken2019,Wright2020,Giblin2021,Hildebrandt2021} to estimate $G_+$ according to Eq.~\eqref{eq:G+def}, with weights $W_{ls} = w_s \Sigma_{\mathrm{crit},ls}^{-2}$ where $w_s$ estimates the precision of the ellipticity measurement of the source $s$ \citep{Brouwer2021,Giblin2021}.
We take into account statistical uncertainties of $G_+$ induced by the statistical uncertainties of the measured tangential ellipticities $\epsilon_+$ following \citet{Brouwer2021,Viola2015},
\begin{align}
\sigma^2_{G_+} (R)
= \frac{
 \sum_s W_{ls}^2 \cdot \Sigma^2_{\mathrm{crit},ls} \sigma^2_{\epsilon_s}
}{
 \left( \sum_s W_{ls} \right)^2
}
\,,
\end{align}
where we follow the notation from Eq.~\eqref{eq:G+def} and where $\sigma_{\epsilon_s}$ is the ellipticity dispersion of the source $s$ whose value we take from Table I of \citet{Giblin2021} according to each source's tomographic redshift bin.

We handle multiplicative and additive biases of the ellipticities as well as their statistical uncertainties as in \citet{Mistele2023d}.
First, to take into account additive biases, we measure the stacked tangential reduced shear around about 45 million uniform random coordinates within the KiDS-DR4 footprint and, for each galaxy cluster under consideration, subtract the result from each cluster's measured $G_+$.
Then, to take into account multiplicative bias, we multiply the measured $G_+$ by a factor of $1/(1+\mu)$ where, following \cite{Brouwer2021}, we adopt $1+\mu = 0.98531$.
The same multiplicative correction is applied to $\sigma_{G_+}$.

To minimize contamination, we apply the $griz$\footnote{
 KiDS provides $gri$ but not $z$.
 To deal with this, we replace $z$ with $Z$, which KiDS does provide, and, for simplicity, ignore the small difference between $z$ and $Z$ \citep{Hewett2006}.
} color-color cuts proposed in \citet[][Table A.1]{EuclidCollaboration2024Lesci} and restrict the sources to those satisfying $z_B > z_l + \Delta z$ where $z_B$ is the photometric redshift of the source, $z_l$ is the redshift of the lens, and $\Delta z = 0.1$.

We calculate the critical surface densities $\Sigma_{\mathrm{crit},ls}$ following the procedure of \citet{Dvornik2017,Brouwer2021} to take into account uncertainties in the source redshifts,
\begin{align}
 \Sigma^{-1}_{\mathrm{crit},ls} = \frac{4 \pi G_{\mathrm{N}}}{c^2} D(z_l) \int_{z_l}^{\infty} d z_s \, n_{ls}(z_s) \cdot \frac{D(z_l, z_s)}{D(z_s)}  \,.
\end{align}
The function $n_{ls}(z)$ is determined as follows.
For a given $z_l$ and a given photometric source redshift $z_{B,s}$, we check in which of the five tomographic bins from \citet{Hildebrandt2021} the $z_{B,s}$ value belongs, get the corresponding redshift distribution function from \citet{Hildebrandt2021}, and normalize this distribution to unity in the interval $[z_l, \infty)$.

We take cluster coordinates and redshifts from the LC2 catalog.
For compatibility with LC2,  we assume a flat FLRW cosmology with $H_0 = 70\,\mathrm{km}\,\mathrm{s}^{-1}\,\mathrm{Mpc}^{-1}$ and $\Omega_m = 0.3$.
We use 10 logarithmic radial bins with the largest bin edge being $3\,\mathrm{Mpc}$ and with a logarithmic bin width of $1/7.5$.
We do not apply any miscentering corrections.

\begin{figure}
\begin{center}
\includegraphics[width=\columnwidth]{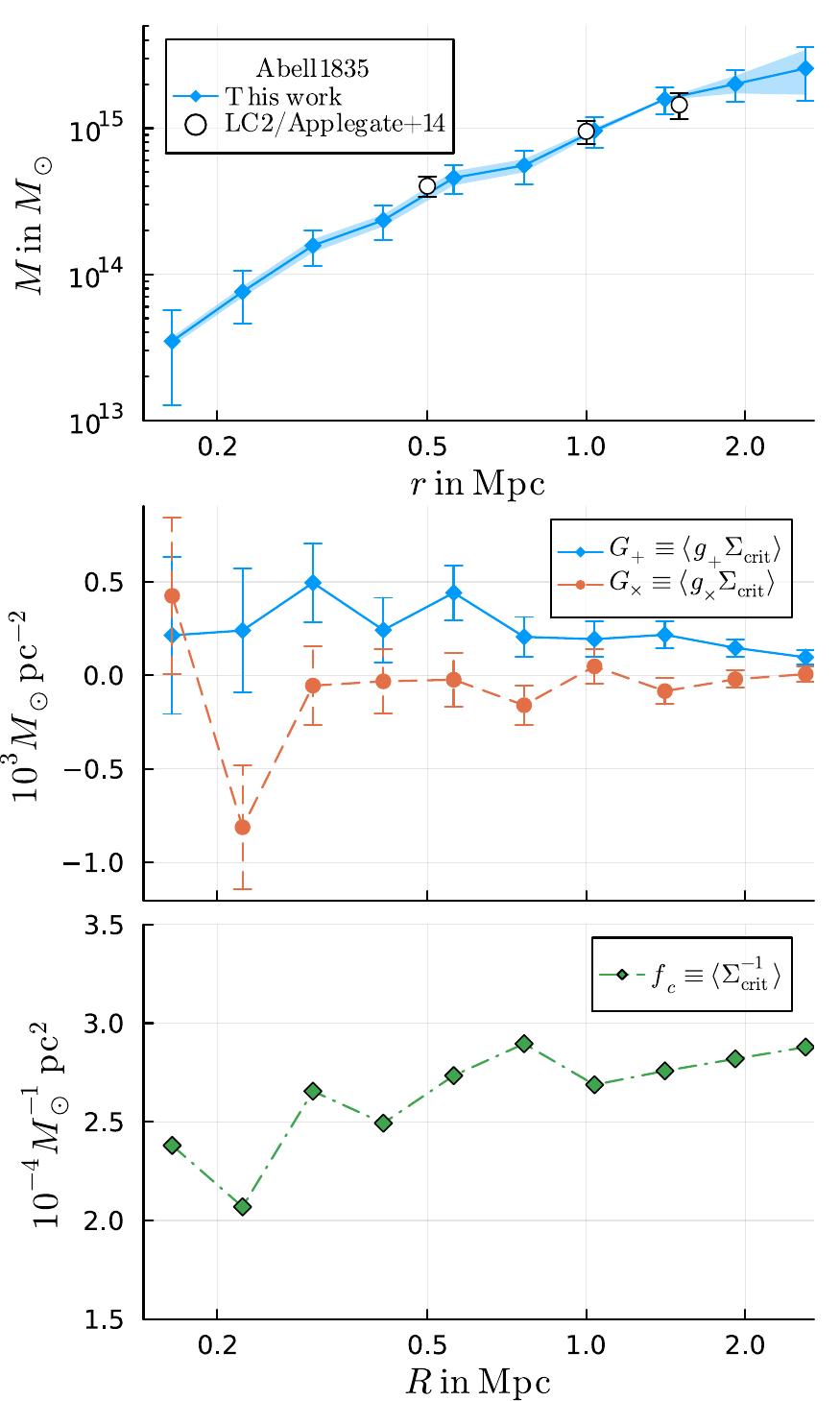}
\end{center}
\caption{
 \emph{Top}: The deprojected spherical mass $M(r)$ of Abell 1835 inferred using our new method described in Sec.~\ref{sec:method} (blue symbols).
 Error bars indicate the statistical uncertainty.
 The colored error band indicates the systematic uncertainties from extrapolating beyond the last data point and interpolating between the discrete data points (see Sec.~\ref{sec:method:practice}).
 For comparison, white symbols show mass estimates from \citet{Applegate2014}, obtained using different data and a different method.
  Our statistical uncertainties are larger because we make fewer assumptions about the mass profile and because the KiDS data we use has a smaller source number density.
 \emph{Middle}: The observed tangential reduced shear $G_+ = \langle g_+ \Sigma_{\mathrm{crit}} \rangle$ and the corresponding cross component $G_\times$.
 The cross component $G_\times$ is consistent with zero, as it should be.
 \emph{Bottom}: The observed average inverse critical surface density $\f = \langle \Sigma_{\mathrm{crit}}^{-1} \rangle$.
 The observational quantities $G_+$, $G_\times$, and $\f$ are inferred from KiDS DR4 data.
 $G_+$ and $\f$ enter the calculation of $M$ through Eq.~\eqref{EQ:ESD_FROM_GF} and Eq.~\eqref{EQ:M_FROM_ESD}.
}
\label{fig:abell1835}
\end{figure}

\begin{figure}
\begin{center}
\includegraphics[width=\columnwidth]{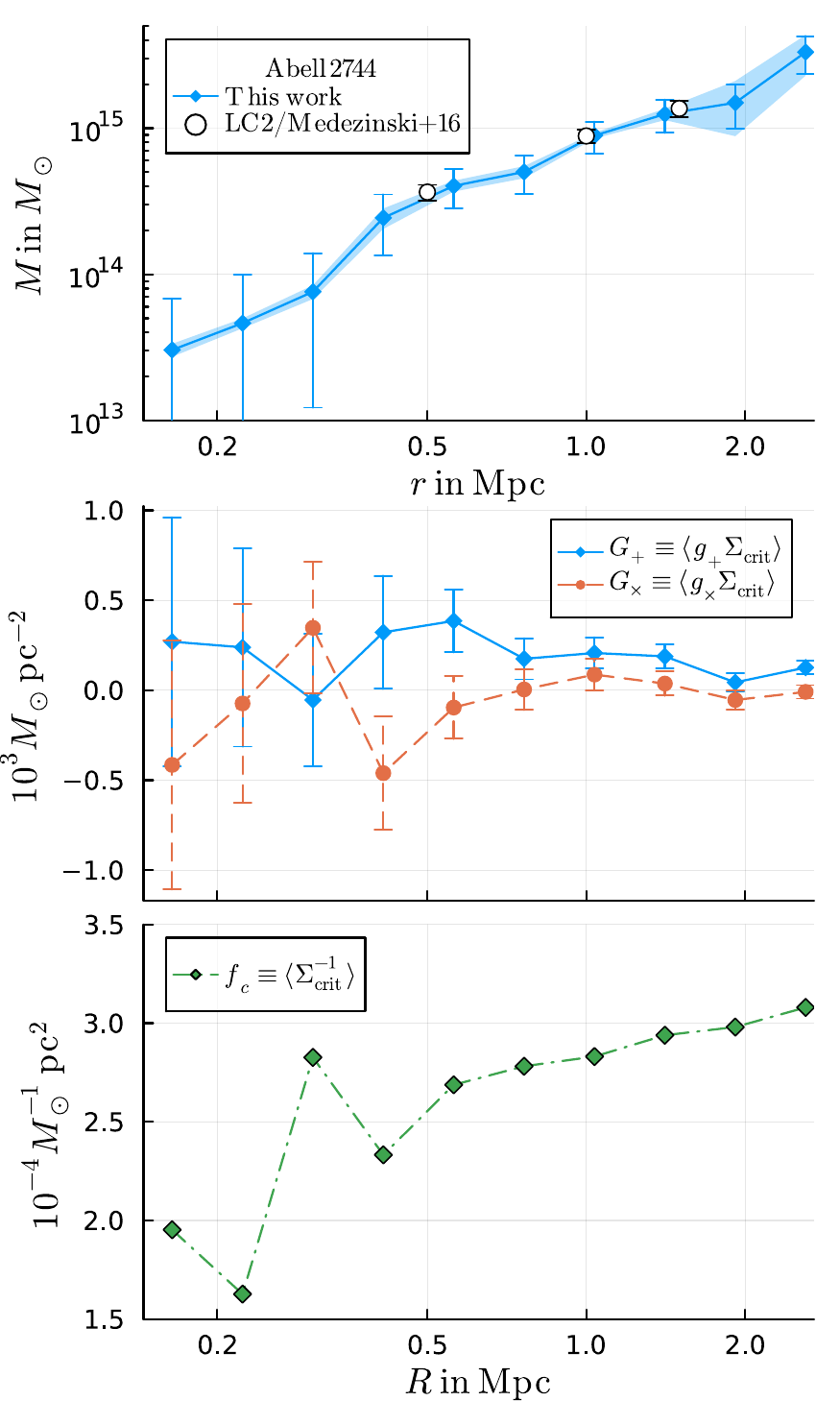}
\end{center}
\caption{Same as Fig.~\ref{fig:abell1835} but for Abell 2744. White symbols in the top panel show mass estimates from \citet{Medezinski2016}.}
\label{fig:abell2744}
\end{figure}

\subsection{Results}

Fig.~\ref{fig:abell1835} and Fig.~\ref{fig:abell2744} show the inferred deprojected mass profiles $M(r)$ (top panels) for Abell 1835 and Abell 2744, respectively, and the observed quantities $G_+(R)$ and $\f(R)$ (middle and bottom panels) from which $M(r)$ is calculated using Eq.~\eqref{EQ:ESD_FROM_GF} and Eq.~\eqref{EQ:M_FROM_ESD}.
As a cross-check, the middle panels also show the cross component of the reduced shear, denoted $G_\times$ and defined analogously to Eq.~\eqref{eq:G+def}, which is consistent with zero, as it should be.

We see that the systematic uncertainties from having to extrapolate and interpolate the observed $G_+(R)$ and $\f(R)$ profiles (see Sec.~\ref{sec:method:practice}) become comparable to the statistical uncertainties only close to the last measured data point.
Over most of radial range we probe, these systematics are relatively unimportant.
Thus, the inferred masses are free from any assumptions about the mass profile of the lens over most of the radial range.

The statistical uncertainties of our mass estimate for Abell 2744 are significantly larger than those of \citet{Medezinski2016}.
This is likely in part due to the fact that we make fewer assumptions about the mass profile of the lens.
In addition, the number density of source galaxies for Abell 2744 in KiDS DR4, about $4.6\,\mathrm{arcmin}^{-2}$, is significantly smaller than the source number density of about $10\,\mathrm{arcmin}^{-2}$ in \citet{Medezinski2016}.
Thus, despite making fewer assumptions, we expect that our method would give uncertainties closer to those of \citet{Medezinski2016} when applied to the same underlying data.
For Abell 1835, our mass uncertainties are only slightly larger than those of \citet{Applegate2014}.
This may be because \citet{Applegate2014} exclude data at radii smaller than $750\,\mathrm{kpc}$ from their fit which significantly increases the statistical uncertainties.

As mentioned above, Abell 2744 is a merging system which is likely not spherically symmetric at small radii.
It may, however, still be approximately spherically symmetric at large radii.
Thus, as argued in Sec.~\ref{sec:method:nonsymm}, our method may still infer the correct mass at large radii, despite the existence of a non-symmetric inner region.
In contrast, a global single-halo fit such as the one from \citet{Medezinski2016} that we compare to, may infer an incorrect mass even at large radii in such situations.
It may therefore be surprising that we find masses consistent with the single-halo fit from \citet{Medezinski2016}.
However, in practice, our statistical uncertainties are likely too large to see a difference.

Indeed, \citet{Medezinski2016} have also done a more involved multi-halo analysis that does not assume spherical symmetry and found that the total $M_{200}$ changes by about $15\%$.
This is smaller than our statistical uncertainties, which are about $25\%-30\%$ in the radial range where Fig.~\ref{fig:abell2744} compares to \citet{Medezinski2016}.
This may explain why our masses agree with the single-halo fit from \citet{Medezinski2016}.

\begin{figure}
\begin{center}
\includegraphics[width=\columnwidth]{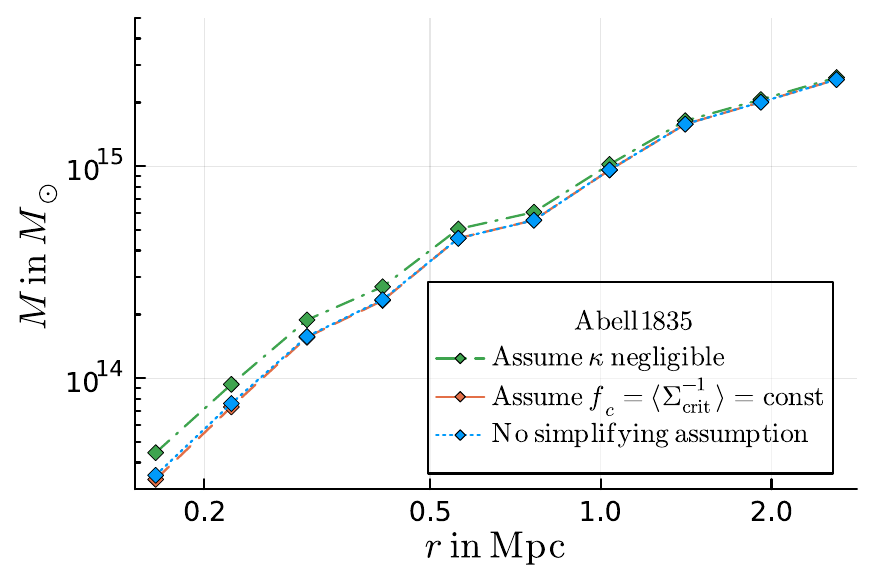}
\end{center}
\caption{
 The inferred deprojected mass $M(r)$ of Abell 1835, as in Fig.~\ref{fig:abell1835}, but making use of additional simplifying assumptions.
 For comparison we also show the original result without these simplifications (dotted blue line).
 The two simplifying assumptions considered are that the average inverse critical surface density $\f = \langle \Sigma_{\mathrm{crit}}^{-1}\rangle$ is approximately constant (dashed red line) and that the convergence $\kappa$ is negligible (dash-dotted green line).
 The former allows using the simpler Eq.~\eqref{eq:ESD_from_Gf_fconst} instead of Eq.~\eqref{EQ:ESD_FROM_GF} to infer $\Delta \Sigma$ from $G_+$ and $\f$.
 The latter corresponds to the even simpler relation $\Delta \Sigma = G_+$.
 For visual clarity, we do not show the uncertainties.
}
\label{fig:methodcompare}
\end{figure}

As discussed in Sec.~\ref{sec:method:theory}, Eq.~\eqref{EQ:ESD_FROM_GF} simplifies if the average inverse critical surface density $\f(R)$ is constant.
Fig.~\ref{fig:abell1835} and Fig.~\ref{fig:abell2744} show that $\f$ is indeed reasonably constant.\footnote{
  Fig.~\ref{fig:abell1835} and Fig.~\ref{fig:abell2744} also show that $\f$ is not \emph{perfectly} constant, with a slight downward trend at small radii.
 One potential cause is residual contamination from cluster members.
 Another is obscuration, which affects high-redshift sources more than low-redshift ones \citep[e.g.][Fig.~7]{Kleinebreil2024}.
 This will be further investigated in future work.
}
Thus, we repeat our analysis for Abell 1835 using the simplified formula Eq.~\eqref{eq:ESD_from_Gf_fconst} which assumes that $\f$ is constant.
In particular, we take the constant value of $\f$ to be the value of $\f(R)$ calculated in a single radial bin that spans the whole radial range we consider.
Fig.~\ref{fig:methodcompare} shows that this does not significantly affect the inferred mass $M(r)$.
Thus, the simplifying assumption of a constant $\f$ may be justified in practice.

The purpose of Eq.~\eqref{EQ:ESD_FROM_GF} (and the simplified Eq.~\eqref{eq:ESD_from_Gf_fconst}) is to take into account that the convergence $\kappa$ is not always negligible in galaxy clusters.
To see this effect of a non-zero $\kappa$, we also repeat our analysis with the assumption that $\kappa$ is negligible, i.e. using $G_+(R) = \Delta \Sigma(R)$ instead of Eq.~\eqref{EQ:ESD_FROM_GF}.
Fig.~\ref{fig:methodcompare} shows that there is a small but significant effect at small and intermediate radii, but not at large radii.
This is expected since $\kappa$ becomes less important at large radii.

\begin{figure}
\begin{center}
\includegraphics[width=\columnwidth]{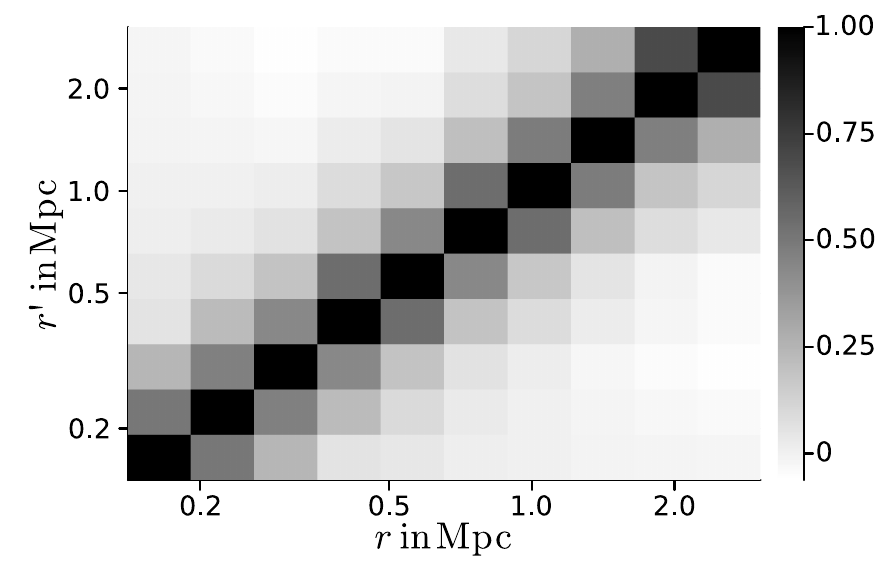}
\end{center}
\caption{
 The correlation matrix of inferred masses $M$ at different radii $r$ and $r'$ for Abell 1835.
 The correlation matrix is defined in terms of the covariance matrix as $\Cov(M(r), M(r'))/\sigma_{M(r)}\sigma_{M(r')}$.
 }
\label{fig:correlationabell1835}
\end{figure}

Due to the integrals in Eq.~\eqref{EQ:ESD_FROM_GF} and Eq.~\eqref{EQ:M_FROM_ESD}, the inferred masses $M(r)$ at different radii $r$ are correlated.
The correlation matrix for Abell 1835 is shown in Fig.~\ref{fig:correlationabell1835}.
We see that the inferred masses are correlated only over a relatively limited radial range.
This is because, as discussed in Sec.~\ref{sec:method:theory}, most of the signal that goes into $M(r)$ comes from $G_+(R)$ and $\f(R)$ at radii $R$ relatively close to $r$.
This is also the reason why the systematic uncertainties from extrapolating beyond the last data point are limited to radii close to the last data point.

\subsection{Future improvements}

The above explicit examples illustrate our new method and demonstrate that it works well.
Further improvements are possible.
For example, the statistical uncertainties can be significantly reduced by using data with higher source number densities.
This may be possible with upcoming surveys such as Euclid \citep{EuclidCollaboration2024c} or dedicated, targeted observations such as those used by \citet{Applegate2014,Medezinski2016} we compared to above.

There are also a number of physical effects we have not yet taken into account.
For example, effects from residual contamination may be relevant at small radii.
At large radii, the lensing signal may receive additional contributions from the correlated environment around the galaxy cluster (i.e. the two-halo term) and the assumption of spherical symmetry may be violated.
One way to take this into account is to estimate the local environment's contribution $\Delta \Sigma_e$ to the ESD profile $\Delta \Sigma$ \citep{Oguri2011b,Oguri2011c} and use this to calculate, using Eq.~\eqref{EQ:M_FROM_ESD}, the local environment's contribution $M_e(r)$ to our inferred mass $M(r)$.
We can then add this contribution $M_e(r)$ to our systematic error estimate.
Alternatively, we can subtract $M_e(r)$ from the inferred mass $M(r)$, at the cost of making the final subtracted mass more model-dependent.

Similarly, the contributions from the uncorrelated large-scale structure can be taken into account as an additive contribution to the covariance matrix of $\Delta \Sigma$ \citep{Hoekstra2003}.
That covariance can be propagated into the final statistical uncertainties and covariance matrices of the mass inferred from $\Delta \Sigma$ using Eq.~\eqref{EQ:M_FROM_ESD}.
Such improvements will be investigated in future work.

\section{Conclusion}
\label{sec:conclusion}

We have introduced a new, non-parametric method to infer galaxy cluster mass profiles from weak lensing, assuming only spherical symmetry.
We have further shown that the assumption of spherical symmetry can be relaxed to allow for a non-symmetric inner region, given a mild additional assumption on the probability distributions of the source redshifts.
We have also discussed how miscentering can be taken into account in a statistical and approximate way and we have demonstrated that this method gives results consistent with other methods and can be computed efficiently.

\section*{Acknowledgements}
We thank Federico Lelli, Stacy McGaugh, and Pengfei Li for helpful comments and discussions.
We thank the anonymous reviewer for comments and suggestions that have significantly improved the manuscript.
This work was supported by the DFG (German Research Foundation) – 514562826.
AD was supported by the European Regional Development Fund and the Czech Ministry of Education, Youth and Sports: Project MSCA Fellowship CZ FZU I -- CZ.02.01.01/00/22\_010/0002906.
Based on observations made with ESO Telescopes at the La Silla Paranal Observatory under programme IDs 177.A-3016, 177.A-3017, 177.A-3018 and 179.A-2004, and on data products produced by the KiDS consortium. The KiDS production team acknowledges support from: Deutsche Forschungsgemeinschaft, ERC, NOVA and NWO-M grants; Target; the University of Padova, and the University Federico II (Naples).

\bibliographystyle{mnras}
\bibliography{cluster-deprojection}

\begin{appendix}

\section{Derivation of Eq.~\eqref{EQ:ESD_FROM_GF} and Eq.~\eqref{EQ:M_FROM_ESD} (Deprojection)}
\label{sec:appendix:derivation}

We start with Eq.~\eqref{eq:GSigmarelation}.
This is a relation between, on the one hand, the quantities $G_+$ and $\f$ that we consider as given and, on the other hand, $\Delta \Sigma$ and $\Sigma$.
The idea is to transform this equation into a linear ordinary differential equation (ODE) for $\Delta \Sigma$ which can be solved analytically.
This will give Eq.~\eqref{EQ:ESD_FROM_GF}.
Having obtained $\Delta \Sigma$ in this way, we can then use previous results from \citet{Mistele2024} to convert $\Delta \Sigma$ into the 3D mass profile $M(r)$.
This will give Eq.~\eqref{EQ:M_FROM_ESD}.

Our first step is to multiply Eq.~\eqref{eq:GSigmarelation} by $1 - \f \Sigma$ and use Eq.~\eqref{eq:ESD_SD_relation} to eliminate $\Sigma$ in favor of $\Delta \Sigma$,
\begin{align}
 G_+(R) \left( 1 + \f(R) \Delta \Sigma(R) - \f(R) \int_R^\infty dR' \frac{2 \Delta \Sigma(R')}{R'} \right) = \Delta \Sigma (R) \,.
\end{align}
Rearranging, we get
\begin{align}
 \label{eq:ESD_equation_before_taking_derivative}
 \Delta \Sigma (R) \left(1 - \frac{1}{G_+ \f(R)}\right) - \int_R^\infty dR' \frac{2 \Delta \Sigma(R')}{R'} = - \frac{1}{\f(R)} \,.
\end{align}
This turns into a linear ODE in $\Delta \Sigma$ after taking a derivative with respect to $R$.
This ODE can be written as
\begin{align}
 \label{eq:ESD_ode}
 \partial_R \Delta \Sigma (R) + \Delta \Sigma (R) \left[\frac{2}{H(R) R} + \partial_R \ln(-H(R))\right] = - \frac{1}{H(R)} \partial_R \frac{1}{\f(R)} \,,
\end{align}
with
\begin{align}
\label{eq:Hdef}
H(R) \equiv 1 - \frac{1}{G_+ \f(R)} \,.
\end{align}
The general solution of Eq.~\eqref{eq:ESD_ode} is
\begin{align}
\label{eq:ESD_solution_initial}
\Delta \Sigma(R) = 
 E(R) \cdot
 \left[
  \frac{C_0}{H(R_0)} + 
  \int_R^{R_0} dR'' \frac{1}{H(R'')} \left(\partial_{R''} \frac{1}{\f(R'')} \right)
  \frac{1}{E(R'')}
 \right] \,,
\end{align}
where $C_0$ is an integration constant, $R_0$ is an auxiliary radius, and the function $E(R)$ is defined as
\begin{align}
E(R) \equiv  \exp\left(\int_R^{R_0} dR' \left(\frac{2}{H(R')R'} + \partial_{R'} \ln(-H(R')) \right) \right) \,.
\end{align}
We can evaluate the integral over the derivative of the logarithm in $E(R)$,
\begin{align}
E(R) = \frac{H(R_0)}{H(R)} \exp\left(
  \int_R^{R_0} \frac{dR'}{R'} \frac{2}{H(R')}
 \right) \,.
\end{align}
Putting this back into Eq.~\eqref{eq:ESD_solution_initial} then gives
\begin{align}
\Delta \Sigma(R) = 
\frac{1}{H(R)} \exp\left(
  \int_R^{R_0} \frac{dR'}{R'} \frac{2}{H(R')}
 \right)  \cdot
 \left[
  C_0 + 
  \int_R^{R_0} dR'' \left(\partial_{R''} \frac{1}{\f(R'')} \right)
  \exp\left(
   -\int_{R''}^{R_0} \frac{dR'}{R'} \frac{2}{H(R')}
  \right)
  \right] \,.
\end{align}
We can replace $R_0$ with $\infty$ by absorbing some constant terms into the integration constant $C_0$,
\begin{align}
\Delta \Sigma(R) = 
\frac{1}{H(R)} \exp\left(
  \int_R^\infty \frac{dR'}{R'} \frac{2}{H(R')}
 \right)  \cdot
 \left[
  C_0 + 
  \int_R^\infty dR'' \left(\partial_{R''} \frac{1}{\f(R'')} \right)
  \exp\left(
   -\int_{R''}^\infty \frac{dR'}{R'} \frac{2}{H(R')}
  \right)
  \right] \,.
\end{align}
This expression contains a derivative of the observational quantity $\f(R)$ which, in practice, can be hard to reliably estimate.
To avoid this numerical derivative, we integrate by parts and absorb some more constants into $C_0$,
\begin{align}
 \label{eq:ESD_before_C0}
\Delta \Sigma(R) = 
\frac{1}{H(R)}  \left[
  C_0 
    \exp\left(
      \int_R^\infty \frac{dR'}{R'} \frac{2}{H(R')}
     \right)
   - \frac{1}{f(R)}
   - \int_R^\infty dR'' \frac{1}{\f(R'')} \frac{2}{R''} \frac{1}{H(R'')}
  \exp\left(
   \int_R^{R''} \frac{dR'}{R'} \frac{2}{H(R')}
  \right)
  \right] \,.
\end{align}

It remains to fix this integration constant.
To do this, we use our assumption that the mass density of the lens falls off faster than $1/r$ at $r \to \infty$.
We also assume that $\f$ stays roughly constant.
These assumptions together with the original Eq.~\eqref{eq:GSigmarelation} imply
\begin{align}
 \lim_{R \to \infty} \frac{\Delta \Sigma(R)}{G_+(R)} = 1\,,
\end{align}
and Eq.~\eqref{eq:ESD_before_C0} then implies
\begin{align}
 C_0 = 0 \,.
\end{align}
Using $C_0 = 0$ and the definition of $H(R)$ from Eq.~\eqref{eq:Hdef} in Eq.~\eqref{eq:ESD_before_C0} then gives the desired result Eq.~\eqref{EQ:ESD_FROM_GF}.

This allows us to infer $\Delta \Sigma(R)$ from the observables $G_+(R)$ and $\f(R)$.
It remains to convert $\Delta \Sigma (R)$ to the deprojected 3D mass $M(r)$.
To this end, we use Eq.~\eqref{EQ:M_FROM_ESD} which was previously derived in Appendix~A of \citet{Mistele2023d}.
For completeness, we here reiterate the main steps of that derivation.
\begin{enumerate}
 \item We first use Eq.~\eqref{eq:ESD_SD_relation} to write $\Sigma$ in terms of $\Delta \Sigma$, i.e.
 \begin{align}
  \label{eq:repeated_ESD_SD_relation}
  \Sigma(R) = - \Delta \Sigma  (R) + \int_R^\infty dR' \frac{2 \Delta \Sigma (R')}{R'} \,.
 \end{align}
 \item We use a deprojection formula from \citet{Kent1986} to convert $\Sigma (R)$ into the 3D mass $M(r)$,
  \begin{align}
   \label{eq:Kent}
  M(r) = \int_0^r dR \, 2 \pi R \, \Sigma(R) + 4 \int_r^\infty dR \, R \, \Sigma(R) \left(\arcsin\left(\frac{r}{R}\right)  - \frac{r}{\sqrt{R^2 -r^2}} \right) \,,
  \end{align}
  which is valid in spherical symmetry.
  This formula can be derived by using an Abel transform that connects the 3D density $\rho$ and the surface density $\Sigma$.
 \item Substituting Eq.~\eqref{eq:repeated_ESD_SD_relation} into Eq.~\eqref{eq:Kent} results in a nested double integral.
       After exchanging the order of the $R$ (from Eq.~\eqref{eq:Kent}) and $R'$ (from Eq.~\eqref{eq:repeated_ESD_SD_relation}) integrals, the inner $R$ integral can be done analytically.
       After some simplifications and relabeling $R' \to R$ in the remaining integral, we end up with
   \begin{align}
    M(r) = 4 r \int_r^\infty dR \, \Delta \Sigma(R) \left(\frac{1}{\sqrt{1 - \left(\frac{r}{R}\right)^2}} - \sqrt{1 - \left(\frac{r}{R}\right)^2}\right) \,.
   \end{align}
 \item The final step is to substitute $r/R = \sin \theta$, which gives Eq.~\eqref{EQ:M_FROM_ESD}, i.e.
  \begin{align}
    M(r) = 4 r^2 \int_0^{\pi/2} d \theta \, \Delta \Sigma \left(\frac{r}{\sin \theta}\right) \,.
  \end{align}

\end{enumerate}

\section{Efficient numerical evaluation}
\label{sec:appendix:numerical}

Using Eq.~\eqref{EQ:M_FROM_ESD} together with Eq.~\eqref{EQ:ESD_FROM_GF} to infer the mass profile $M(r)$ from the observational quantities $G_+(R)$ and $\f(R)$ means evaluating a nested triple integral for each $r$.
Such nested integrals can be expensive to evaluate numerically.
Here, we show how to efficiently evaluate this nested triple integral by solving two ODEs and then, for each $r$, evaluating only a single integral.
Each of these steps can be computed efficiently by standard numerical techniques.
To numerically solve ODEs and evaluate integrals we use the Julia packages `OrdinaryDiffEq.jl` \citep{Rackauckas2017} and `QuadGK.jl`, respectively.

We start by defining the integral
\begin{align}
 I(R) \equiv \int_R^\infty dR' \frac{2}{R'} \frac{G_+ \f(R')}{1 - G_+ \f(R')} \,.
\end{align}
This integral can be calculated at each $R$ by solving the ODE
\begin{align}
 \label{eq:Iprime}
 I'(R) = - \frac{2}{R} \frac{G_+\f(R)}{1 - G_+\f(R)} \,,
\end{align}
with the boundary condition
\begin{align}
 I(R_{\mathrm{max}})
  = \int_{R_{\mathrm{max}}}^\infty dR' \frac{2}{R'} \frac{G_+ \f(R')}{1 - G_+ \f(R')} 
  = - \frac2n \ln(1 - G_+\f(R_{\mathrm{max}}))
 \,,
\end{align}
which we evaluated analytically using our extrapolation beyond the last data point from Eq.~\eqref{eq:extrapolate}.
Having solved the ODE for $I(R)$, we can write Eq.~\eqref{EQ:ESD_FROM_GF} as
\begin{align}
 \label{eq:ESD_from_J_and_I}
 \Delta \Sigma (R)
 =
 \frac{1}{\f(R)} 
 \frac{G_+ \f(R)}{1 - G_+ \f(R)}
 \left[
 1
 - e^{-I(R)} \f(R) J(R)
  \right]
  \,,
\end{align}
where $J(R)$ is the integral
\begin{align}
 J(R) \equiv \int_R^\infty dR'' \frac{1}{\f(R'')} \frac{2}{R''} \frac{G_+\f(R'')}{1 - G_+\f(R'')} e^{+I(R'')} \,.
\end{align}
We can evaluate $J(R)$ analogously to $I(R)$.
That is, we solve the ODE
\begin{align}
 \label{eq:Jprime}
J'(R) = - \frac{1}{\f(R)} \frac{2}{R} \frac{G_+\f(R)}{1 - G_+\f(R)} e^{+I(R)} \,,
\end{align}
with boundary condition
\begin{align}
 J(R_{\mathrm{max}})
 = \int_{R_{\mathrm{max}}}^\infty dR'' \frac{1}{\f(R'')} \frac{2}{R''} \frac{G_+\f(R'')}{1 - G_+\f(R'')} e^{+I(R'')}
 = \frac{1}{\f(R_{\mathrm{max}})} \left((1 - G_+\f(R_{\mathrm{max}}))^{-\frac2n} - 1\right)
 \,.
\end{align}
Having solved the ODE for $J(R)$, we can directly obtain $\Delta \Sigma (R)$ from Eq.~\eqref{eq:ESD_from_J_and_I}.
It then only remains to calculate $M(r)$ using Eq.~\eqref{EQ:M_FROM_ESD}, i.e. using
\begin{align}
 \frac{M(r)}{r^2}
   = 4 \int_0^{\pi/2} d \theta \, \Delta \Sigma \left( \frac{r}{\sin \theta}\right)
   \,.
\end{align}
We evaluate this integral numerically with $\Delta \Sigma$ given by Eq.~\eqref{eq:ESD_from_J_and_I}.
For $r/\sin \theta$ smaller than $R_{\mathrm{max}}$, we use the numerical ODE solutions discussed above for $I(R)$ and $J(R)$ in Eq.~\eqref{eq:ESD_from_J_and_I}.
For $r/\sin \theta$ larger than $R_{\mathrm{max}}$, we evaluate $I(R)$ and $J(R)$ analytically using our extrapolation from Eq.~\eqref{eq:extrapolate}. 

As discussed in Sec.~\ref{sec:method:practice}, one may choose to interpolate the discrete $G_+$ and $\f$ data points in either $R$ or $\ln R$.
In practice, the numerical procedure discussed so far is likely most suitable when interpolating in $R$.
When interpolating in $\ln R$, it may be more appropriate to consider the integrals $I$ and $J$ as functions of $\ln R$ instead of $R$ and calculate them by solving ODEs written in terms of $d I / d \ln R$ and $d J / d \ln R$ (instead of Eq.~\eqref{eq:Iprime} and Eq.~\eqref{eq:Jprime} which are written in terms of $dI/dR$ and $dJ/dR$).
It is straightforward to adapt the above procedure to this case by using $d / d \ln R = R \, d/dR$. 
The provided Julia code supports interpolation in both $R$ and $\ln R$ and automatically uses the corresponding way of calculating $I$ and $J$.

\section{Derivation of Eq.~\eqref{EQ:GSIGMARELATION_NONSYMM} (Deprojection with non-symmetric inner region)}
\label{sec:appendix:nonsymm}

Consider a mass distribution that is spherically symmetric only beyond some spherical radius $r_s$ (see Sec.~\ref{sec:method:nonsymm}).
Here, we derive an analog of the relation Eq.~\eqref{eq:GSigmarelation} between $G_+$, $\f$, $\Sigma$, and $\Delta \Sigma$ that holds even with this weaker symmetry assumption.

Consider the azimuthally averaged tangential shear $G_+ = \langle g_+ \cdot \Sigma_{\mathrm{crit}} \rangle$.
The expectation value $\langle \dots \rangle$ here can be understood as, first, separately at each position $(R \cos \varphi, R \sin \varphi)$, taking the expectation value of $g_+ \Sigma_{\mathrm{crit}}$ with respect to the source redshift probability distribution at that position, and then averaging azimuthally over $\varphi$.
That is,
\begin{align}
\begin{split}
 G_+ (R)
  &= \Bigl< \frac{\gamma_+ \Sigma_{\mathrm{crit},ls}}{1 - \kappa} \Bigr> (R)
  = \frac{1}{2\pi} \int_0^{2\pi} d \varphi \int d z_s p_{z_s}(z_s|R) \, \frac{\gamma_+ \Sigma_{\mathrm{crit},ls}}{1 - \Sigma \cdot \Sigma_{\mathrm{crit},ls}^{-1}}  \\
  &= \int d z_s p_{z_s}(z_s|R) \, \frac{1}{2\pi} \int_0^{2\pi} d \varphi \, \frac{\gamma_+ \Sigma_{\mathrm{crit},ls}}{1 - \Sigma \cdot \Sigma_{\mathrm{crit},ls}^{-1}} 
  \,.
\end{split}
\end{align}
Here, the notation $p_{z_s}(z_s|R)$ indicates our assumption that the sources at each position are drawn from a probability distribution that does not depend on the azimuth $\varphi$ (see Sec.~\ref{sec:method:nonsymm}).
This $\varphi$-independence allows us to move the $z_s$ integral including the factor of $p_{z_s}$ outside the $\varphi$ integral (see the second line).
We now follow steps analogous to those in the derivation of Eq.~\eqref{eq:GSigmarelation} \citep[e.g.,][]{Umetsu2020}.
First, we expand for small $\kappa$,
\begin{subequations}
\begin{align}
G_+(R)
  &= \int d z_s p_{z_s}(z_s|R) \, \frac{1}{2\pi} \int_0^{2\pi} d \varphi \, (\gamma_+ \Sigma_{\mathrm{crit},ls}) \cdot (1 + \Sigma \cdot \Sigma_{\mathrm{crit},ls}^{-1} + \dots) \\
  \label{eq:Gexpansion}
  &= \Delta \Sigma (R) + \int d z_s p_{z_s}(z_s|R) \, \Sigma_{\mathrm{crit},ls}^{-1} \, \frac{1}{2\pi} \int_0^{2\pi} d \varphi \, (\gamma_+ \Sigma_{\mathrm{crit},ls}) \cdot \Sigma + \dots \,,
\end{align}
\end{subequations}
where we used that the azimuthal average of $\gamma_+ \cdot \Sigma_{\mathrm{crit}}$ gives the ESD profile $\Delta \Sigma$ even without having to assume spherical symmetry \citep{Kaiser1995,Bartelmann1995}.
One only has to generalize the definition of $\Delta \Sigma$ from Eq.~\eqref{eq:ESD_def} by replacing $\Sigma$ by its azimuthal average $\langle \Sigma \rangle$ (as in Eq.~\eqref{eq:ESD_def_nonsymm_maintext}),
\begin{align}
 \label{eq:ESD_def_nonsymm}
 \Delta \Sigma(R) \equiv \frac{2}{R^2} \int_0^R dR' R' \langle \Sigma \rangle(R') - \langle \Sigma \rangle (R) \,.
\end{align}
Up to this point, we have not made use of any symmetry assumption about the lens, we only made use of $p_{z_s}(z_s|R)$ being independent of $\varphi$.
Thus, we will be able to reuse these formulas (but not what follows) in Appendix~\ref{sec:appendix:miscentering:pR_miscentered_frame} below.

Consider now projected radii $R > r_s$.
Because the density $\rho$ is spherically symmetric beyond spherical radii $r_s$, the surface density $\Sigma$ is cylindrically symmetric at $R > r_s$, i.e. $\Sigma = \Sigma(R)$ at $R > r_s$.
Thus, in the second term of Eq.~\eqref{eq:Gexpansion}, we can pull $\Sigma$ out of the $\varphi$ integral,
\begin{align}
 G_+(R)
 = \Delta \Sigma (R) + \Sigma(R) \int d z_s p_{z_s}(z_s|R) \, \Sigma_{\mathrm{crit},ls}^{-1} \, \frac{1}{2\pi} \int_0^{2\pi} d \varphi \, (\gamma_+ \Sigma_{\mathrm{crit},ls}) + \dots \quad (\mathrm{for}\;R > r_s) \,.
\end{align}
This leaves another azimuthal average of $\gamma_+ \Sigma_{\mathrm{crit}}$ which, as discussed above, we can replace by $\Delta \Sigma$,
\begin{align}
\label{eq:Gsigmarelation_nonsymm_appendix}
G_+(R)
  = \Delta \Sigma (R) + \Delta \Sigma (R) \langle \Sigma_{\mathrm{crit},ls}^{-1} \rangle (R) \, \Sigma (R) + \dots 
  \approx \frac{\Delta \Sigma (R)}{1 - \Sigma (R) \, \f (R)} \quad (\mathrm{for}\;R > r_s) \,,
\end{align}
where we have used the definition $\f = \langle \Sigma_{\mathrm{crit}}^{-1} \rangle$ and have rewritten the Taylor expansion in $\kappa$ as a Padé approximation in the usual way \citep{Seitz1997}.
This is the desired result Eq.~\eqref{EQ:GSIGMARELATION_NONSYMM}.

The above derivation assumes that redshift estimates for individual sources are available.
An analogous argument shows that the result Eq.~\eqref{eq:Gsigmarelation_nonsymm_appendix} holds even when only ensemble information is available.
We then just need to use the alternative definitions of $G_+$ and $\f$ from Eq.~\eqref{eq:G_and_fc_def_ensemble}.

\section{Derivation of Eq.~\eqref{EQ:MISCENTERCORRECT} (Miscentering correction)}
\label{sec:appendix:miscentering}

Consider a spherically symmetric lens with surface density $\Sigma_0$, excess surface density $\Delta \Sigma_0$, azimuthally averaged tangential reduced shear $G_{+,0}$, azimuthally averaged inverse critical surface density $\fzero$, and a 3D mass profile $M_0$.
Here, the subscript ``$0$" indicates that these quantities are defined in the original, centered frame.
In Sec.~\ref{sec:method:theory}, we discuss how to use observations of $G_{+,0}$ and $\fzero$ to infer $M_0$.
Here, we show how to (approximately, at large radii) infer $M_0$ if we observe this mass distribution in a frame that is miscentered by an offset $\Rmc$ and an angle $\theta_{\mathrm{mc}}$.
Due to the miscentering, the mass distribution  does not appear spherically symmetric in the reference frame used for observations.
For example, at a position $(x, y) = (R \cos \varphi, R \sin \varphi)$ in the miscentered frame the surface density $\Sigma(x, y)$ is
\begin{align}
\begin{split}
 \Sigma(R \cos \varphi, R \sin \varphi)
 &= \Sigma_0(\sqrt{ (R \cos \varphi - \Rmc \cos \theta_{\mathrm{mc}})^2 + (R \sin \varphi - \Rmc \sin \theta_{\mathrm{mc}})^2 }) \\
 &= \Sigma_0(\sqrt{R^2 + \Rmc^2 - 2 R \Rmc \cos(\varphi - \theta_{\mathrm{mc}})})\,.
 \end{split}
\end{align}
For later reference, we note that expanding $\Sigma$ to quadratic order in $\epsilon = \Rmc/R$ gives
\begin{multline}
\Sigma(R \cos \varphi, R \sin \varphi)
= \Sigma_0(R) \\
  - \frac{\Rmc}{R} \cos(\varphi - \theta_{\mathrm{mc}}) R \partial_R \, \Sigma_0(R)
  + \frac12 \left(\frac{\Rmc}{R}\right)^2 \left(
    \sin^2(\varphi - \theta_{\mathrm{mc}}) R \partial_R  \Sigma_0(R)
    + \cos^2(\varphi - \theta_{\mathrm{mc}}) R^2 \partial_R^2 \Sigma_0(R)
   \right) \,,
\end{multline}
and that its azimuthal average to the same order is
\begin{align}
\label{eq:Sigma_azimuth_expand}
\frac{1}{2\pi} \int_0^{2\pi} d\varphi \, \Sigma(R \cos \varphi, R \sin \varphi)
= \Sigma_0(R) +  \frac14 \left(\frac{\Rmc}{R}\right)^2 ( R \partial_R  \Sigma_0(R) + R^2 \partial_R^2 \Sigma_0(R)) \,.
\end{align}

Our overall strategy in the following will be to first express the $G_{+,0}$ and $\fzero$ of the original centered lens through the $G_+$ and $\f$ as observed in the miscentered frame by Taylor expanding in $\epsilon = R_{\mathrm{mc}}/R$.
We can then apply our standard deprojection procedure from Sec.~\ref{sec:method:theory} to the $G_{+,0}$ and $\fzero$ obtained in this way to infer the desired mass profile $M_0$ of the original, centered lens.
We will find that both $G_{+,0}$ and $\fzero$ obtain corrections starting at order $\epsilon^2$.
Schematically,
\begin{align}
\begin{split}
 G_{+,0} (R) &= G_+(R) + \epsilon^2 \cdot (\mathrm{terms\;involving\;}G_+{\mathrm{\;and\;its\;derivatives}}) \,, \\
 \fzero  (R) &= \f (R) + \epsilon^2 \cdot (\mathrm{terms\;involving\;}\f{\mathrm{\;and\;its\;derivatives}}) \,.
\end{split}
\end{align}
In practice, the $O(\epsilon^2)$ corrections to $\fzero = \f$ are not needed for our deprojection method.
Indeed, their effect on the final, reconstructed mass will be of order $\kappa \epsilon^2$ which we choose to neglect (see Sec.~\ref{sec:method:miscentering}).
To see this, we note that $\f$ enters our deprojection method as the prefactor of $\Sigma$ in Eq.~\eqref{eq:GSigmarelation} (see Sec.~\ref{sec:method:theory}).
Thus, for our purposes, we can use
\begin{align}
\begin{split}
 G_{+,0} (R) &= G_+(R) + \epsilon^2 \cdot (\mathrm{terms\;involving\;}G_+{\mathrm{\;and\;its\;derivatives}}) \,, \\
 \fzero  (R) &= \f (R) \,.
\end{split}
\end{align}
In the following, our task will therefore be to (i) explicitly calculate the miscentering corrections to $G_{+,0}$ to order $\epsilon^2$ (but we neglect terms of order $\kappa \epsilon^2$) and (ii) check that there are no corrections to $\fzero$ of order $\epsilon$ (but we do not need to explicitly calculate the nonzero corrections of order $\epsilon^2$).

We assume that the source redshifts are drawn from a probability distribution $p_{z_s}$ that does not depend on the azimuth $\varphi$ (see Sec.~\ref{sec:method:miscentering}).
Specifically, we will consider two variations of this assumption: That this azimuth-independence holds in the miscentered frame (Appendix~\ref{sec:appendix:miscentering:pR_miscentered_frame}) and that it holds in the original, centered frame (Appendix~\ref{sec:appendix:miscentering:pR_original_frame}).
As we will show, the result is the same in both cases.

\subsection{Derivation assuming $p_{z_s}$ is azimuth-independent in miscentered frame}
\label{sec:appendix:miscentering:pR_miscentered_frame}

Consider the azimuthally averaged tangential shear $G_+ = \langle g_+ \cdot \Sigma_{\mathrm{crit}} \rangle$ in the miscentered frame.
Here, this frame is also where we assume $p_{z_s}$ to be independent of the azimuth $\varphi$.
Thus, as discussed in Appendix~\ref{sec:appendix:nonsymm}, after expanding in $\kappa$ and without making any symmetry assumptions about the lens, we can write $G_+$ as (see Eq.~\eqref{eq:Gexpansion})
\begin{align}
\label{eq:Gexpansion_repeated_at_miscentering}
G_+(R) = \Delta \Sigma (R) + \int d z_s p_{z_s}(z_s|R) \, \Sigma_{\mathrm{crit},ls}^{-1} \, \frac{1}{2\pi} \int_0^{2\pi} d \varphi \, (\gamma_+ \Sigma_{\mathrm{crit},ls}) \cdot \Sigma + \dots \,,
\end{align}
where $\Delta \Sigma$ is defined as in Eq.~\eqref{eq:ESD_def_nonsymm}.
Expanding the second term in Eq.~\eqref{eq:Gexpansion_repeated_at_miscentering} in $\epsilon = R_{\mathrm{mc}}/R$, we find by explicit calculation that all terms linear in $\epsilon$ are proportional to $\cos(\varphi-\theta_{\mathrm{mc}})$ and therefore vanish after the azimuthal average.
This includes $O(\epsilon)$ terms obtained by expanding $\Sigma$ and $O(\epsilon)$ terms obtained by expanding $\gamma_+ \Sigma_{\mathrm{crit},ls}$.
The terms quadratic in $\epsilon$ in the second term of Eq.~\eqref{eq:Gexpansion_repeated_at_miscentering} do not vanish but are multiplied by $\Sigma$ and so are of order $\kappa \epsilon^2$ which we choose to neglect (see Sec.~\ref{sec:method:miscentering}).
Thus, all corrections due to miscentering in the second term in Eq.~\eqref{eq:Gexpansion_repeated_at_miscentering} are of higher order than our approximation.
We can therefore replace $\Sigma$ in this term by $\Sigma_0$ and pull it out of the $\varphi$ integral because $\Sigma_0$ is independent of $\varphi$.
This leaves an azimuthal average of $\gamma_+ \Sigma_{\mathrm{crit}}$ which, as discussed in Appendix~\ref{sec:appendix:nonsymm}, gives $\Delta \Sigma$.
Thus, we have

\begin{align}
\label{eq:Gexpansion2}
G_+(R)
  = \Delta \Sigma (R) + \Delta \Sigma (R) \langle \Sigma_{\mathrm{crit},ls}^{-1} \rangle (R) \, \Sigma_0 (R) + \dots 
  \approx \frac{\Delta \Sigma (R)}{1 - \Sigma_0 (R) \, \f(R)} \,,
\end{align}
where we have used the definition of $\f (R) = \langle \Sigma_{\mathrm{crit}}^{-1} \rangle (R)$ and we have rewritten the Taylor expansion in $\kappa$ as a Padé approximation in the usual way \citep{Seitz1997}.

We next expand $\Delta \Sigma$ in $\epsilon$.
To this end, we substitute the expanded azimuthal average of $\Sigma$ from Eq.~\eqref{eq:Sigma_azimuth_expand} into the definition of $\Delta \Sigma$ Eq.~\eqref{eq:ESD_def_nonsymm}.
We find up to terms of higher order than $\epsilon^2$,
\begin{align}
 \Delta \Sigma(R) = \Delta \Sigma_0(R) + \frac14 \left(\frac{\Rmc}{R}\right)^2 ( R \partial_R  \Sigma_0(R) - R^2 \partial_R^2 \Sigma_0(R)) \,.
\end{align}
We then use Eq.~\eqref{eq:ESD_SD_relation} to eliminate $\Sigma_0$ in favor of $\Delta \Sigma_0$.
This equation Eq.~\eqref{eq:ESD_SD_relation} applies because $\Sigma_0$ and $\Delta \Sigma_0$ refer to a spherically symmetric mass distribution.
It can be written as $\partial_R \Sigma_0 = - \partial_R \Delta \Sigma_0 - 2 \Delta \Sigma_0 / R$.
We find
\begin{align}
 \Delta \Sigma(R) = \Delta \Sigma_0(R) - \frac14 \left(\frac{\Rmc}{R}\right)^2 (4 \Delta \Sigma_0(R) - R \partial_R \Delta \Sigma_0(R) - R^2 \partial_R^2 \Delta \Sigma_0(R)) \,.
\end{align}
We can plug this result back into Eq.~\eqref{eq:Gexpansion2} and find
\begin{align}
G_+ (R) \approx G_{+,0}(R) - \frac14 \left(\frac{\Rmc}{R}\right)^2 (4 G_{+,0}(R) - R \partial_R G_{+,0}(R) - R^2 \partial_R^2 G_{+,0}(R)) \,,
\end{align}
which holds up to terms of order $\kappa \epsilon^2$ that we neglect.
In particular, we can replace $\f$ by $\fzero$ in the denominator since $\f$ is multiplied by $\Sigma$ and, as we will show below, the difference between $\f$ and $\fzero$ is of order $\epsilon^2$.
Further, since $\Delta \Sigma_0 = G_{+,0} (1 + \cal{O}(\kappa))$ we can replace $\Delta \Sigma_0$ by $G_{+,0}$ in terms multiplied by $\epsilon^2$, and similarly we can neglect the denominator in Eq.~\eqref{eq:Gexpansion2} for terms multiplied by $\epsilon^2$.
Inverting this result to order $\epsilon^2$, we find
\begin{align}
\label{eq:Gexpansion3}
G_{+,0} (R) \approx G_+(R) + \frac14 \left(\frac{\Rmc}{R}\right)^2 (4 G_+(R) - R \partial_R G_+(R) - R^2 \partial_R^2 G_+(R)) \,,
\end{align}
which is the desired result for $G_{+,0}$.

Consider then the azimuthally averaged inverse critical surface density $\f$ in the miscentered frame.
We have
\begin{align}
\begin{split}
\f(R)
&= \langle \Sigma_{\mathrm{crit},ls}^{-1} \rangle(R) \\
&= \frac{1}{2\pi} \int_0^{2 \pi} d \varphi \int d z_s \, p_{z_s}(z_s|R) \, \Sigma_{\mathrm{crit},ls}^{-1} \\
&= \frac{1}{2\pi} \int_0^{2 \pi} d \varphi \int d z_s \, \left[ p_{z_s}(z_s|R) + \frac{R_\mathrm{mc}}{R} \cos(\varphi - \theta_{\mathrm{mc}}) \, R \partial_R p_{z_s}(z_s|R) \right] \, \Sigma_{\mathrm{crit},ls}^{-1} \\
&= \frac{1}{2\pi} \int_0^{2 \pi} d \varphi \int d z_s \, \left[ p_{z_s}\left(z_s|\sqrt{ (R \cos \varphi + \Rmc \cos \theta_{\mathrm{mc}})^2 + (R \sin \varphi + \Rmc \sin \theta_{\mathrm{mc}})^2 }\right) + O(\epsilon^2) \right] \, \Sigma_{\mathrm{crit},ls}^{-1} \,,
\end{split}
\end{align}
where we used that the azimuthal average of $\cos(\varphi - \theta_{\mathrm{mc}})$ vanishes.
The first term in the last line is just $\fzero(R)$ observed in the original, centered frame.
Thus, we have,
\begin{align}
 \fzero(R) = \f(R) + O(\epsilon^2)\,,
\end{align}
which is the desired result for $\fzero$.

\subsection{Derivation assuming $p_{z_s}$ is azimuth-independent in original, centered frame}
\label{sec:appendix:miscentering:pR_original_frame}

It remains to show that the same results for $G_+$ and $\f$ hold when assuming that $p_{z_s}$ is independent of the azimuth in the original, centered frame rather than in the miscentered frame.
We first consider $G_+$ as measured in the miscentered frame and follow the same first few steps as in Appendix~\ref{sec:appendix:nonsymm}, but we take into account that $p_{z_s}$ is not $\varphi$-independent in this frame,
\begin{align}
\label{eq:Gsigmaexpansion_pR_originalframe}
\begin{split}
 G_+ (R)
  &= \Bigl< \frac{\gamma_+ \Sigma_{\mathrm{crit},ls}}{1 - \kappa} \Bigr> (R) 
  = \frac{1}{2\pi} \int_0^{2\pi} d \varphi \int d z_s p_{z_s}^{\mathrm{mc}}(z_s|R, \varphi) \, \frac{\gamma_+ \Sigma_{\mathrm{crit},ls}}{1 - \Sigma \cdot \Sigma_{\mathrm{crit},ls}^{-1}} \\
  &= \frac{1}{2\pi} \int_0^{2\pi} d \varphi \int d z_s p_{z_s}^{\mathrm{mc}}(z_s|R,\varphi) \, (\gamma_+ \Sigma_{\mathrm{crit},ls})(1 + \Sigma \cdot \Sigma_{\mathrm{crit},ls}^{-1} + \dots) \\
  &= \Delta \Sigma + \frac{1}{2\pi} \int_0^{2\pi} d \varphi \int d z_s p_{z_s}^{\mathrm{mc}}(z_s|R,\varphi) \, (\gamma_+ \Sigma_{\mathrm{crit},ls}) \Sigma \cdot \Sigma_{\mathrm{crit},ls}^{-1} + \dots   \,,
\end{split}
\end{align}
where we used the shorthand notation
\begin{align}
p_{z_s}^{\mathrm{mc}}(z_s|R, \varphi) = 
p_{z_s}\left(z_s|\sqrt{(R \cos \varphi - R_{\mathrm{mc}} \cos \theta_{\mathrm{mc}})^2 + (R \sin \varphi - R_{\mathrm{mc}} \sin \theta_{\mathrm{mc}})^2}\right)\,.
\end{align}
Moving the $z_s$ integral including the factor of $p^{\mathrm{mc}}_{z_s}$ out of the $\varphi$ integral is, at this point, not possible due to the $\varphi$-dependence of $p^{\mathrm{mc}}_{z_s}(z_s|R,\varphi)$.
For the first term, which gives $\Delta \Sigma$, we used that the combination $\gamma_+ \Sigma_{\mathrm{crit},ls}$ is a property of only the lens and is independent of $z_s$ (since $\gamma$ depends on $z_s$ only through an overall $1/\Sigma_{\mathrm{crit}}$ scaling).
This allows to pull the $\gamma_+ \Sigma_{\mathrm{crit},ls}$ factor out of the $z_s$ integral, after which the $z_s$ integral of this first term is just $1$ because $p^{\mathrm{mc}}_{z_s}$ is normalized to $1$.

When expanding in $\epsilon$, the second term in Eq.~\eqref{eq:Gsigmaexpansion_pR_originalframe} contains one more $O(\epsilon)$ term compared to Eq.~\eqref{eq:Gexpansion_repeated_at_miscentering} because we also have to expand $p_{z_s}^{\mathrm{mc}}$.
This term is, however, again proportional to $\cos(\varphi - \theta_{\mathrm{mc}})$ which again vanishes due to the azimuthal average.
Thus, as in Eq.~\eqref{eq:Gexpansion_repeated_at_miscentering}, miscentering corrections in the second term in  Eq.~\eqref{eq:Gsigmaexpansion_pR_originalframe} start at order $\kappa \epsilon^2$ which we neglect.
Thus, we can replace $p^{\mathrm{mc}}_{z_s}$ and $\Sigma$ by their zeroth-order expressions in this second term,
\begin{align}
 G_+ (R)
  &= \Delta \Sigma + \frac{1}{2\pi} \int_0^{2\pi} d \varphi \int d z_s p_{z_s}(z_s|R) \, (\gamma_+ \Sigma_{\mathrm{crit},ls}) \Sigma_0 \cdot \Sigma_{\mathrm{crit},ls}^{-1} + \dots
  \,.
\end{align}
We can then swap the $\varphi$ and $z_s$ integrals and pull $p_{z_s}$, $\Sigma_0$, and $\Sigma_{\mathrm{crit},ls}^{-1}$ out of the $\varphi$ integral to obtain
\begin{align}
G_+(R)
  = \Delta \Sigma (R) + \Delta \Sigma (R) \fzero(R)  \, \Sigma_0 (R) + \dots 
  \approx \frac{\Delta \Sigma (R)}{1 - \Sigma_0 (R) \, \fzero(R)} \,.
\end{align}
Compared to Eq.~\eqref{eq:Gexpansion2}, the denominator now has $\fzero$ instead of $\f$.
Indeed, due to the different assumption about $p_{z_s}$, we now have that $\int dz_s p_{z_s}(z_s|R) \Sigma_{\mathrm{crit},ls}^{-1}$ is $\fzero(R)$ instead of $\f(R)$.
From here on we can follow the same steps as in Appendix~\ref{sec:appendix:miscentering:pR_miscentered_frame} to arrive at the final result for $G_+$ Eq.~\eqref{eq:Gexpansion3}.

For the azimuthally averaged inverse critical surface density $\f$ in the miscentered frame, we have
\begin{align}
\begin{split}
\f(R)
&= \langle \Sigma_{\mathrm{crit},ls}^{-1} \rangle(R) \\
&= \frac{1}{2\pi} \int_0^{2 \pi} d \varphi \int d z_s \, p_{z_s}^{\mathrm{mc}}(z_s|R, \varphi) \, \Sigma_{\mathrm{crit},ls}^{-1} \\
&= \frac{1}{2\pi} \int_0^{2 \pi} d \varphi \int d z_s \, \left[ p_{z_s}(z_s|R) - \frac{R_\mathrm{mc}}{R} \cos(\varphi - \theta_{\mathrm{mc}}) \, R \partial_R p_{z_s}(z_s|R) + O(\epsilon^2) \right] \, \Sigma_{\mathrm{crit},ls}^{-1} \\
&= \frac{1}{2\pi} \int_0^{2 \pi} d \varphi \int d z_s \, p_{z_s}(z_s|R) \, \Sigma_{\mathrm{crit},ls}^{-1} + O(\epsilon^2) \\
&= \fzero(R) + O(\epsilon^2) \,.
\end{split}
\end{align}
This is the desired result for $\fzero$.

The above derivations assume that redshift estimates for individual sources are available.
Analogous arguments show that the results hold even when only ensemble information is available.
We then just need to use the alternative definitions of $G_+$ and $\f$ from Eq.~\eqref{eq:G_and_fc_def_ensemble}.

\end{appendix}

\end{document}